\documentclass[12pt]{article}
\pdfoutput=1
\usepackage{color}
\usepackage{amssymb}
\usepackage{epsfig}

\usepackage{graphicx}

\def\L{\mathcal L}
\def\e{\varepsilon}

\newcommand{\wt}{\widetilde}

\begin{document}

\def\a{\alpha}
\def\b{\beta}
\def\c{\chi}
\def\d{\delta}
\def\e{\epsilon}
\def\f{\phi}
\def\g{\gamma}
\def\h{\eta}
\def\i{\iota}
\def\j{\psi}
\def\k{\kappa}
\def\l{\lambda}
\def\m{\mu}
\def\n{\nu}
\def\o{\omega}
\def\p{\pi}
\def\q{\theta}
\def\r{\rho}
\def\s{\sigma}
\def\t{\tau}
\def\u{\upsilon}
\def\x{\xi}
\def\z{\zeta}
\def\D{\Delta}
\def\F{\Phi}
\def\G{\Gamma}
\def\J{\Psi}
\def\L{\Lambda}
\def\O{\Omega}
\def\P{\Pi}
\def\Q{\Theta}
\def\S{\Sigma}
\def\U{\Upsilon}
\def\X{\Xi}

\def\ve{\varepsilon}
\def\vf{\varphi}
\def\vr{\varrho}
\def\vs{\varsigma}
\def\vq{\vartheta}

\def\dg{\dagger}                                     
\def\ddg{\ddagger}                                   
\def\wt#1{\widetilde{#1}}                    
\def\mt{\widetilde{m}_1}
\def\mti{\widetilde{m}_i}
\def\rt{\widetilde{r}_1}
\def\mtt{\widetilde{m}_2}
\def\mttt{\widetilde{m}_3}
\def\rtt{\widetilde{r}_2}
\def\mb{\overline{m}}
\def\VEV#1{\left\langle #1\right\rangle}        
\def\be{\begin{equation}}
\def\ee{\end{equation}}
\def\ds{\displaystyle}
\def\ra{\rightarrow}

\def\bea{\begin{eqnarray}}
\def\eea{\end{eqnarray}}
\def\NO{\nonumber}
\def\Bar#1{\overline{#1}}


\def\pl#1#2#3{Phys.~Lett.~{\bf B {#1}} ({#2}) #3}
\def\np#1#2#3{Nucl.~Phys.~{\bf B {#1}} ({#2}) #3}
\def\prl#1#2#3{Phys.~Rev.~Lett.~{\bf #1} ({#2}) #3}
\def\pr#1#2#3{Phys.~Rev.~{\bf D {#1}} ({#2}) #3}
\def\zp#1#2#3{Z.~Phys.~{\bf C {#1}} ({#2}) #3}
\def\cqg#1#2#3{Class.~and Quantum Grav.~{\bf {#1}} ({#2}) #3}
\def\cmp#1#2#3{Commun.~Math.~Phys.~{\bf {#1}} ({#2}) #3}
\def\jmp#1#2#3{J.~Math.~Phys.~{\bf {#1}} ({#2}) #3}
\def\ap#1#2#3{Ann.~of Phys.~{\bf {#1}} ({#2}) #3}
\def\prep#1#2#3{Phys.~Rep.~{\bf {#1}C} ({#2}) #3}
\def\ptp#1#2#3{Progr.~Theor.~Phys.~{\bf {#1}} ({#2}) #3}
\def\ijmp#1#2#3{Int.~J.~Mod.~Phys.~{\bf A {#1}} ({#2}) #3}
\def\mpl#1#2#3{Mod.~Phys.~Lett.~{\bf A {#1}} ({#2}) #3}
\def\nc#1#2#3{Nuovo Cim.~{\bf {#1}} ({#2}) #3}
\def\ibid#1#2#3{{\it ibid.}~{\bf {#1}} ({#2}) #3}

\title{
\vspace*{1mm}
\bf Strong thermal $SO(10)$-inspired leptogenesis in the light of 
recent results from long-baseline neutrino experiments}
\author{
{\Large Marco Chianese and Pasquale Di Bari }
\\
{\it Physics and Astronomy}, 
{\it University of Southampton,} \\
{\it  Southampton, SO17 1BJ, U.K.}
}

\maketitle \thispagestyle{empty}

\vspace{-10mm}

\begin{abstract}
We confront recent experimental results on neutrino mixing parameters with the requirements from strong thermal $SO(10)$-inspired leptogenesis, where the asymmetry is produced from  next-to-lightest right-handed neutrinos $N_2$ independently of the initial conditions. There is a nice agreement with  latest global analyses supporting  $\sin\delta < 0$  and normal ordering at $ \sim 95\%$ C.L. On the other hand, the more stringent experimental lower bound on the atmospheric mixing angle starts to corner strong thermal $SO(10)$-inspired leptogenesis. Prompted and encouraged by this rapid experimental advance, we obtain a precise determination of the allowed region in the plane $\delta$ versus $\theta_{23}$. We confirm that for the benchmark case $\a_2 \equiv m_{D2} / m_{\rm charm}= 5 \, $, where $m_{D2}$ is the intermediate neutrino Dirac mass setting the $N_2$ mass, and initial pre-existing  asymmetry $N_{B-L}^{\rm p, i} = 10^{-3}$, the bulk of  solutions lies in the first octant. Though most of the solutions are found outside the $95\%$ C.L. experimental region, there is still a big allowed fraction that does not require a too fine-tuned choice of the Majorana phases so that the neutrinoless double beta decay effective neutrino mass allowed range is still $m_{ee}\simeq [10,30]\,{\rm meV}$. We also show how the constraints depend on  $N_{B-L}^{\rm p, i}$   and $\alpha_2$. In particular, 
we show that the current best fit, ($\theta_{23},\delta)\simeq (47^{\circ}, -130^{\circ})$, can be reproduced
for $N_{B-L}^{\rm p, i} = 10^{-3}$ and $\alpha_2 = 6$. Such large  values for $\a_2$ have been
recently obtained in a few  realistic fits within $SO(10)$-inspired models. Finally, we also obtain that current 
neutrino data rule out $N_{B-L}^{\rm p, i} \gtrsim 0.1$ for $\alpha_2 \lesssim 4.7$.
\end{abstract}

\newpage
\section{Introduction}

In the absence of clear signs of new physics  at the TeV scale or below, 
it is reasonable that  an explanation of neutrino masses and mixing
is associated to the existence of higher energy scales.   In particular, 
a conventional high energy type I seesaw mechanism \cite{seesaw}
can also account for the matter-antimatter asymmetry of the Universe,
via high energy scale leptogenesis \cite{fy}.
This is currently regarded as the most minimal and attractive  possibility. 

Latest global analyses from neutrino oscillation experiments 
also  seem to rule out $C\!P$ conservation in left-handed (LH) neutrino mixing 
at $90\% \, {\rm C.L.}$ \cite{nufit} (see also  \cite{marrone} and \cite{valencia} for previous analyses).
This is not a sufficient condition for the existence of a source of $C\!P$
violation for successful leptogenesis but, if confirmed,
it would be still an important result because it would make reasonable to have
$C\!P$ violation also in heavy right-handed (RH) neutrinos, 
the dominant source of $C\!P$
violation for leptogenesis (excluding special scenarios) while  $C\!P$ 
conservation in LH neutrino mixing  could legitimately rise doubts on it.
Moreover, the exclusion of quasi-degenerate  light neutrino masses
is also a positive experimental result for minimal leptogenesis scenarios, 
based on high scale type I seesaw mechanism and thermal RH neutrino production, 
since these typically require values of neutrino masses $m_i \lesssim {\cal O}(0.1)\,{\rm eV}$ \cite{upperbound,bounds},  even taking into account charged lepton \cite{flavour} and heavy neutrino \cite{geometry} flavour effects.
Therefore, the current phenomenological picture encourages
the investigation of high energy scale scenarios of leptogenesis.  

The possibility to test leptogenesis in a statistically significant way
relies on the identification of specific scenarios, 
possibly emerging from well motivated theoretical frameworks. 
This should increase the predictive power to a level that the seesaw parameter
space can be over-constrained and that the probability 
that the predictions are just a mere coincidence becomes very low. 
Though this strategy is certainly challenging, it received an important support by the 
 measurement of a value of the reactor mixing angle  sufficiently large to allow a completion of  the measurements of the  unknown parameters in the leptonic mixing matrix:   $C\!P$ violating Dirac phase, 
neutrino mass ordering and a determination of the deviation of the atmospheric mixing angle 
from its maximal value.

The latest results from the NO$\nu$A  \cite{NOVA} and T2K \cite{T2K} long baseline 
neutrino experiments seem to exclude  a deviation 
of the atmospheric mixing angle from maximal
mixing larger than $\sim 5^{\circ}$
and support negative values of $\sin\d$. 
They also show an emerging  preference for normally ordered neutrino masses (NO)
compared to inverted ordered neutrino masses (IO). 
When all results are combined, a recent global analysis
finds that  NO is preferred  at $\sim 2\,\s$ \cite{nufit}. 
Moreover it is found that the best fit occurs for the atmospheric mixing angle in the second octant,
though  first octant is disfavoured only very slightly, at less than $\simeq 0.5\,\s$, an important
point for our study.

This emerging  experimental set of results for the unknown neutrino oscillation parameters 
is potentially in agreement with the expectations from the so-called strong thermal $SO(10)$-inspired leptogenesis (STSO10) solution \cite{strongSO10} requiring NO and approximately negative 
$\sin\d$. However,  for the wash-out of large values of an initial pre-existing asymmetry 
$N_{B-L}^{\rm p,i} \gtrsim 10^{-3}$ and for $\a_2 \equiv m_{D2} / m_{\rm charm}  \lesssim 5$,
the STSO10 also requires the atmospheric mixing angle $\theta_{23}$ to lie in the first octant,
a result that we will further confirm in our analysis though with our improved numerical
procedure we could find  marginal solutions for $\theta_{23}$ as large as $45.75^{\circ}$.
Therefore, a measurement of the atmospheric mixing angle in the second octant 
would basically rule out the STSO10 solution for $N_{B-L}^{\rm p,i} \gtrsim 10^{-3}$ 
and $\a_2 \lesssim 5$.

The STSO10 is based on two   independent conditions and it is non trivial 
that they can be satisfied simultaneously. 
The first condition, from a model building perspective, is the {\em $SO(10)$-inspired condition} \cite{SO10inspired}.
It corresponds to assume that the  Dirac neutrino mass matrix is {\em not too different} from the up-quark mass
matrix, a typical feature of different grand-unified models (not only $SO(10)$ models). 
The second condition, on the other hand, is purely cosmological, 
requiring that the final asymmetry not only reproduces the observed 
one ({\em successful leptogenesis condition}) but also, less trivially, 
that is independent of  the initial conditions (the {\em strong thermal leptogenesis condition}).
In particular this implies that a possible large pre-existing initial $B-L$ asymmetry is efficiently washed-out.\footnote{This condition is well motivated by the fact that at the large required values of the reheat temperatures,
$T_{RH} \gtrsim 10^{10}\,{\rm GeV}$, different mechanisms can produce asymmetries
much larger than the observed ones. In particular within GUT models, the decays of different
particles heavier than the RH neutrinos, such as heavy gauge bosons, can also produce a sizeable 
asymmetry (thermally or even non thermally).}
For hierarchical RH neutrino mass patterns, the strong thermal condition 
is satisfied only for a seemingly very special case: the tauon $N_2$-dominated scenario \cite{problem}. 
It is then intriguing that, imposing $SO(10)$-inspired conditions,
one finds  a subset of solutions also satisfying independence of the initial conditions, 
thus realising the STSO10 solution.

The full allowed region in the plane $\d$ versus $\theta_{23}$ requested by the STSO10 solution
has not been yet firmly determined. 
At the large values of $\theta_{23}$ allowed by latest experimental results, the  range of $\delta$ from the STSO10 gets much narrower than
the current experimental $2\,\sigma$ interval given  by $\d\simeq [-190^{\circ},-30^{\circ}]$ \cite{nufit}. 
Therefore, a precise determination, both theoretical and experimental, of the 
region $\delta$ versus $\theta_{23}$  can provide a powerful test of STSO10.

It should be clearly said that all constraints from the STSO10 solution depend on the value of the initial pre-existing 
$B-L$ asymmetry $N_{B-L}^{\rm p,i}$ to be washed-out.  The higher is 
the value of $N_{B-L}^{\rm p,i}$, the most stringent the constraints are and there is
a maximum value of $N_{B-L}^{\rm p,i}$ above which there is no allowed region.

The goal of this paper is to determine precisely, within the given set of assumptions
and approximations,  the allowed window for $\d$ as a function of $\theta_{23}$, and at the same time 
the upper bound on $\theta_{23}$ for a given value of $N_{B-L}^{\rm p,i}$  and $\alpha_2$. 
As in previous papers \cite{strongSO10,
SO10decription,VLimpact}, we use as benchmark values $N_{B-L}^{\rm p,i} =10^{-3}$ and $\alpha_2=5$.
For this case we double check the constraints  comparing results obtained numerically diagonalising
the inverse Majorana mass matrix in the Yukawa basis with those obtained 
using the analytical procedure presented in \cite{SO10decription}
and extended in \cite{VLimpact} taking into account the mismatch between the Yukawa basis and the weak basis,
since this helps enhancing the asymmetry and consequently enlarging the allowed window on $\d$. 
This further supports the validity of the analytic procedure\footnote{Notice that in \cite{SO10decription,VLimpact} the comparison between analytical and numerical procedures 
has been done comparing the calculation of $C\!P$ asymmetries, baryon asymmetry 
and flavoured decay parameters versus $m_1$ for a selected set of 
benchmark solutions. However, the  different constraints on low energy neutrino parameters from 
$SO(10)$-inspired leptogenesis were still derived numerically for the general case $I \leq V_L \leq V_{CKM}$ 
and the allowed region from STSO10 was obtained extracting, from the solutions satisfying successful $SO(10)$-inspired leptogenesis,
that subset  also satisfying  the strong thermal condition. 
This should make clear the difference between the results obtained in this paper with those
obtained in \cite{SO10decription,VLimpact}.}
that is then used to derive, in a much more efficient way,
the dependence of the constraints not only on $N_{B-L}^{\rm p,i}$ but also on the other theoretical parameter
$\a_2 \equiv m_{D2}/m_{\rm charm}$. We should stress that in this paper
we manage for the first time to saturate the allowed region on $\delta$ versus $\theta_{23}$ in  the STSO10 solution
thanks to the generation a much higher number of solutions (${\cal O}(10^6)$), about three orders of magnitude more, 
compared to previous analyses \cite{strongSO10,SO10decription,VLimpact}. This has been possible
by virtue mainly of two reasons: first, here we focus just on the STSO10 solution, while in previous analyses this 
was extracted as a subset from the more general set of $SO(10)$-inspired leptogenesis solutions; 
second, the use of the analytical procedure
found in \cite{SO10decription,VLimpact} avoids the lengthy diagonalisation 
of the inverse Majorana mass matrix in the Yukawa basis
leading to a much faster generation of solutions. 
As we said, however, for the benchmark case the constraints were crossed checked 
also using the usual numerical procedure.  

The paper is organised as follows. In Section 2 we review the seesaw type I mechanism and current neutrino oscillation data. 
In Section 3 we review briefly the STSO10 leptogenesis solution and the analytical procedure that we follow for the derivation of the results.
In Section 4  we determine the allowed region in the plane $\delta$ versus $\theta_{23}$ for the benchmark case
$N_{B-L}^{\rm p,i}=10^{-3}$ and $\a_2=5$. In Section 5 we show the dependence of the  constraints on $N_{B-L}^{\rm p,i}$ and $\a_2$.
For $\a_2 =6$ the allowed region gets significantly enhanced in the plane $\delta$ versus $\theta_{23}$, allowing  the 
current best fit values  $\theta_{23} \simeq 47^{\circ}$ and $\delta \simeq -130^{\circ}$ though
at the expense of a fine-tuning in the Majorana phases. In Section 6 we draw conclusions.
 
\section{Seesaw and low energy neutrino parameters}

Augmenting the SM with three RH neutrinos $N_{R i}$
with Yukawa couplings $h$ and a Majorana mass term M, in the flavour basis where both
the charged lepton mass matrix and $M$ are diagonal, one can write   
the leptonic mass terms  generated after spontaneous symmetry breaking 
by the Higgs expectation value $v$ as ($\a=e,\m,\t$ and $i=1,2,3$)
\be
- {\cal L}_{M} = \,  \overline{\a_L} \, D_{m_{\ell}}\,\a_R + 
                              \overline{\nu_{\a L}}\,m_{D\a i} \, N_{R i} +
                               {1\over 2} \, \overline{N^{c}_{R i}} \, D_{M} \, N_{R i}  + \mbox{\rm h.c.}\, ,
\ee
where $D_{m_{\ell}} \equiv {\rm diag}(m_e,m_{\m},m_{\t})$, $D_{M}\equiv {\rm diag}(M_1,M_2,M_3)$
and $m_{D}=h\, v$ is the neutrino Dirac mass matrix.  
In the seesaw limit, for $M \gg m_D$, the mass spectrum splits into two sets of Majorana eigenstates,
a light set with masses $m_{1} \leq m_{2} \leq m_3$ given by the seesaw formula \cite{seesaw}
\be\label{seesaw}
D_m =  U^{\dagger} \, m_D \, {1\over D_M} \, m_D^T  \, U^{\star}  \,   ,
\ee
with $D_m = {\rm diag}(m_1,m_2,m_3)$, and a heavy set with masses basically
coinciding with the three $M_i$'s in $D_M$. The matrix $U$, diagonalising the light neutrino mass
matrix $m_{\nu} = -m_D\,M^{-1}\,m_D^T$ in the weak basis, has then to be identified
with the PMNS lepton mixing matrix. 

For NO, the PMNS matrix can be parameterised
in terms of the usual mixing angles $\theta_{ij}$, the Dirac phase $\d$
and the Majorana phases $\rho$ and $\s$, as
\be
U=  \left( \begin{array}{ccc}
c_{12}\,c_{13} & s_{12}\,c_{13} & s_{13}\,e^{-{\rm i}\,\d} \\
-s_{12}\,c_{23}-c_{12}\,s_{23}\,s_{13}\,e^{{\rm i}\,\d} &
c_{12}\,c_{23}-s_{12}\,s_{23}\,s_{13}\,e^{{\rm i}\,\d} & s_{23}\,c_{13} \\
s_{12}\,s_{23}-c_{12}\,c_{23}\,s_{13}\,e^{{\rm i}\,\d}
& -c_{12}\,s_{23}-s_{12}\,c_{23}\,s_{13}\,e^{{\rm i}\,\d}  &
c_{23}\,c_{13}
\end{array}\right)
\, {\rm diag}\left(e^{i\,\rho}, 1, e^{i\,\sigma}
\right)\,   .
\ee
Since the STSO10 solution cannot be realised for IO, we can focus on NO. 
In this case latest neutrino oscillation experiments global analyses find for the mixing angles and the 
leptonic Dirac phase $\d$ the following best fit values , $1\s$ errors  and $3\s$ intervals \cite{nufit}: 
\bea\label{expranges}
\theta_{13} & = &  8.54^{\circ}\pm 0.15^{\circ} \in [8.09^{\circ}, 8.98^{\circ}] \,  , \\ \nonumber
\theta_{12} & = &  33.62^{\circ}\pm 0.77^{\circ} \in [31.42^{\circ}, 36.05^{\circ}]  \,  , \\ \nonumber
\theta_{23} & = &  {47.2^{\circ}}^{+1.9^{\circ}}_{-3.9} \in [40.3^{\circ}, 51.5^{\circ}]  \,  ,  \\ \nonumber
\d & = &  {-126^{\circ}} ^{+43^{\circ}}_{-31^{\circ}} \in  [-216^{\circ}, 14^{\circ}]  \, .
\eea 
Interestingly there is already a $3\s$ exclusion interval, $\d\ni [14^\circ, 144^{\circ}]$
and $\sin\d > 0$ is excluded at about $2\s$ favouring  $\sin\d < 0$, while,
on the other hand, there are no  experimental constraints on the Majorana phases so far. 
Neutrino oscillation experiments are also sensitive to the differences of squared neutrino masses, 
finding  for the solar neutrino mass scale 
$m_{\rm sol}\equiv \sqrt{m^{\, 2}_2 - m_1^{\, 2}} = (8.6\pm 0.1)\,{\rm meV}$
and for the atmospheric neutrino mass scale 
$m_{\rm atm}\equiv \sqrt{m^{\, 2}_3 - m_1^{\, 2}} = (49.9\pm 0.3)\,{\rm meV}$.

No signal of neutrinoless double beta ($0\nu\b\b$) decays has been detected and, therefore, 
experiments place an upper bound on the effective 
$0\nu\b\b$ neutrino mass $m_{ee} \equiv |m_{\nu ee}|$. 
The most stringent one, so far, has been set by the KamLAND-Zen collaboration 
that found $m_{ee} \leq (61 \mbox{---} 165)\,{\rm meV}$  ($90\%\, {\rm C.L.}$) \cite{kamlandzen},
where the range accounts for nuclear matrix element uncertainties. 

Finally, cosmological observations are sensitive to the 
sum of  neutrino masses. The {\em Planck} satellite collaboration 
 placed a robust stringent upper bound $\sum_i m_i \lesssim 170\,{\rm meV}$ at $95\% {\rm C.L.}$
\cite{planck16}. Once experimental values
of the solar and atmospheric neutrino mass scales are taken into account, 
this translates into an upper bound on the 
lightest neutrino mass $m_1 \lesssim 50\,{\rm meV}$.

\section{Strong thermal $SO(10)$-inspired leptogenesis}

Let us now briefly review  strong thermal $SO(10)$-inspired leptogenesis.
The neutrino Dirac mass matrix can be diagonalised (singular value decomposition or bi-unitary
parameterisation) as
\be\label{svd}
m_D = V^{\dagger}_L \, D_{m_D} \, U_R \,  ,
\ee
where $D_{m_D} \equiv {\rm diag}(m_{D1},m_{D2},m_{D3})$ and where
$V_L$ and $U_R$ are two unitary matrices acting respectively on the
LH and RH neutrino fields and operating the transformation from the 
weak basis (where $m_{\ell}$ is diagonal) to the Yukawa basis (where $m_D$ is diagonal).  

If we parameterise the neutrino Dirac masses $m_{Di}$ in terms of the up quark masses,\footnote{For the values of the 
up-quark masses at the scale of leptogenesis, we adopt  
$(m_{\rm up},m_{\rm charm}, m_{\rm top})=(1\,{\rm MeV}, 400\,{\rm MeV}, 100\,{\rm GeV})$ \cite{charm}.}
\be
(m_{D 1}, m_{D 2}, m_{D 3})=(\a_1 \, m_{\rm up}, \a_2\, m_{\rm charm}, \a_3 \, m_{\rm top})  \,  ,
\ee
we impose {\em $SO(10)$-inspired conditions} \cite{SO10inspired,afs,riotto1} defined as
\begin{itemize}
\item $\a_i = {\cal O}(0.1$--$10)$ \,  ;
\item $I\leq V_L \lesssim V_{CKM}$ \,   .
\end{itemize}
With the latter we imply that  parameterising $V_L$
in the same way as the leptonic mixing matrix $U$, the three
mixing angles $\theta_{12}^L$, $\theta_{23}^L$ and $\theta_{13}^L$
do not have values  much larger than the three mixing angles
in the CKM matrix and in particular 
$\theta_{12}^L \lesssim \theta_{c} \simeq 13^{\circ}$,
where $\theta_c$ is the Cabibbo angle.\footnote{More precisely we adopt: $\theta_{12}^L \leq  13^{\circ} \simeq \theta_{12}^{CKM} \equiv \theta_c$, 
$\theta_{23}^L \leq  2.4^{\circ}\simeq \theta_{23}^{CKM}$, $\theta_{13}^L \leq  0.2^{\circ}\simeq \theta_{13}^{CKM}$.}

Rewriting the seesaw formula Eq.~(\ref{seesaw})  by means of the singular value decomposed 
form Eq.~(\ref{svd}) for $m_D$, one obtains
\be\label{invM}
M^{-1} \equiv  U_R \, D_M \, U_R^T = - D_{m_D}^{-1} \, \widetilde{m}_{\nu} \, D_{m_D}^{-1} \,  ,
\ee
where $M \equiv U^{\star}_R\,D_M\,U^{\dagger}_R$ and 
$\widetilde{m}_{\nu} \equiv V_L\,m_{\nu}\,V_L^T$ 
are respectively  the Majorana mass matrix and  the light neutrino mass matrix 
in the Yukawa basis.  Diagonalising the matrix on the RH side of Eq.~(\ref{invM}),
one can express the RH neutrino masses and the RH neutrino mixing matrix $U_R$
in terms of $m_{\nu}$, $V_L$ and the three $\a_i$'s. 

From the analytical procedure discussed in \cite{afs,SO10decription,VLimpact},
one finds simple expressions for the three RH neutrino masses, 
\be\label{RHspectrum}
M_1    \simeq    {\a_1^2 \,m^2_{\rm up} \over |(\widetilde{m}_\nu)_{11}|} \, , \;\;
M_2  \simeq     {\a_2^2 \, m^2_{\rm charm} \over m_1 \, m_2 \, m_3 } \, {|(\widetilde{m}_{\nu})_{11}| \over |(\widetilde{m}_{\nu}^{-1})_{33}|  } \,  ,  \;\;
M_3  \simeq   \a_3^2\, {m^2_{\rm top}}\,|(\widetilde{m}_{\nu}^{-1})_{33}| \, ,
\ee
and for the RH neutrino mixing matrix
\be
U_R \simeq  
\left( \begin{array}{ccc}
1 & -{m_{D1}\over m_{D2}} \,  {\widetilde{m}^\star_{\nu 1 2 }\over \widetilde{m}^\star_{\nu 11}}  & 
{m_{D1}\over m_{D3}}\,
{ (\widetilde{m}_{\n}^{-1})^{\star}_{13}\over (\widetilde{m}_{\n}^{-1})^{\star}_{33} }   \\
{m_{D1}\over m_{D2}} \,  {\widetilde{m}_{\nu 12}\over \widetilde{m}_{\nu 11}} & 1 & 
{m_{D2}\over m_{D3}}\, 
{(\widetilde{m}_{\n}^{-1})_{23}^{\star} \over (\widetilde{m}_{\n}^{-1})_{33}^{\star}}  \\
 {m_{D1}\over m_{D3}}\,{\widetilde{m}_{\nu 13}\over \widetilde{m}_{\nu 11}}  & 
- {m_{D2}\over m_{D3}}\, 
 {(\widetilde{m}_\nu^{-1})_{23}\over (\widetilde{m}_\nu^{-1})_{33}} 
  & 1 
\end{array}\right) 
\,  D_{\Phi} \,  ,
\ee
with the three phases in 
$D_{\phi} \equiv {\rm diag}(e^{-i \, {\Phi_1 \over 2}}, e^{-i{\Phi_2 \over 2}}, e^{-i{\Phi_3 \over 2}})$ 
given by \cite{VLimpact}
\bea
\Phi_1 & =  & {\rm Arg}[-\widetilde{m}_{\nu 11}^{\star}] \,  , \\
\Phi_2 & = & {\rm Arg}\left[{\widetilde{m}_{\nu 11}\over (\widetilde{m}_{\nu}^{-1})_{33}}\right] -2\,(\rho+\s) - 2\,(\rho_L + \s_L) \, , \\
\Phi_3 & =  & {\rm Arg}[-(\widetilde{m}_{\nu}^{-1})_{33}] \,  .
\eea
One can also derive an  expression for the 
orthogonal matrix starting from its definition
$\O = D_m^{-{1\over 2}}\, U^{\dagger} \, m_D \,  D_M^{-{1\over 2}} $ \cite{casas}
that, using Eq.~(\ref{svd}), becomes \cite{riotto1}
\be
\O= D_m^{-{1\over 2}}\, U^{\dagger} \, V_L^{\dagger} \, D_{m_D} \, U_R \, D_M^{-{1\over 2}} \,  ,
\ee
or in terms of its matrix elements 
\be\label{Oij}
\O_{ij} \simeq {1\over \sqrt{m_i \, M_j}} \, 
\sum_k \, m_{D l} \, U^{\star}_{ki}\,V^{\star}_{L\,lk}\,U_{R \, k j} \,  ,
\ee
from which one finds \cite{VLimpact}
\be\label{Omegaapp}
\hspace{-9mm}\O \simeq 
\left( \begin{array}{ccc}
i\, {(\widetilde{m}_{\nu}\,W^{\star})_{11} \over \sqrt{m_1 \, \widetilde{m}_{\nu 11}}} & 
\sqrt{m_2\,m_3\,(\widetilde{m}_{\nu}^{-1})_{33} \over \widetilde{m}_{\nu 11}}\,
\left(W^{\star}_{2 1} - W^{\star}_{31}\,{{(\widetilde{m}_{\nu}^{-1})_{23}}\over (\widetilde{m}_{\nu}^{-1})_{33}}\right) & 
{W^{\star}_{31}\over \sqrt{m_1\,(\widetilde{m}_{\nu}^{-1})_{33}}} \\
i\, {(\widetilde{m}_{\nu}\,W^{\star})_{12} \over \sqrt{m_2 \, \widetilde{m}_{\nu 11}}} & 
\sqrt{m_1\,m_3\,(\widetilde{m}_{\nu}^{-1})_{33} \over \widetilde{m}_{\nu 11}}\,
\left(W^{\star}_{22} - W^{\star}_{32}\,{{(\widetilde{m}_{\nu}^{-1})_{23}}\over (\widetilde{m}_{\nu}^{-1})_{33}}\right)  
& {W^{\star}_{32}\over \sqrt{m_2\,(\widetilde{m}_{\nu}^{-1})_{33}}}  \\
i\, {(\widetilde{m}_{\nu}\,W^{\star})_{13} \over \sqrt{m_3 \, \widetilde{m}_{\nu 11}}}  & 
\sqrt{m_1\,m_2\,(\widetilde{m}_{\nu}^{-1})_{33} \over \widetilde{m}_{\nu 11}}\,
\left(W^{\star}_{2 3} - W^{\star}_{3 3}\,{{(\widetilde{m}_{\nu}^{-1})_{23}}\over (\widetilde{m}_{\nu}^{-1})_{33}}\right)  
& {W^{\star}_{33}\over \sqrt{m_3\,(\widetilde{m}_{\nu}^{-1})_{33}}}  
\end{array}\right)  \,   ,
\ee
where we introduced $W \equiv V_L \, U$.

Let us now discuss the calculation of the matter-antimatter asymmetry of the universe within leptogenesis.
This can be expressed in terms of the baryon-to-photon number ratio, whose measured value 
from {\em Planck} data (including lensing) combined with
external data sets \cite{planck15}, is found
\be\label{etaBexp}
\eta_{B}^{\rm exp} = (6.10 \pm 0.04)\, \times 10^{-10}  \,   . 
\ee
  We are interested in those solutions satisfying at the same time successful leptogenesis and strong thermal condition. If in general we assume that the  final asymmetry is given by the sum of two terms,
\be\label{2terms}
N_{B-L}^{\rm f} = N_{B-L}^{\rm p,f} + N_{\rm B-L}^{\rm lep,f}  \,  ,
\ee
where the first term is the relic value of a pre-existing asymmetry
and the second is the asymmetry generated from leptogenesis,
the baryon-to-photon number ratio is then also given by 
the sum of two contributions, $\eta_B^{\rm p}$ and $\eta_B^{\rm lep}$,
respectively. The typical assumption 
is that the initial pre-existing asymmetry, after inflation and prior to leptogenesis, is negligible.
Suppose that some external mechanism has generated
a large value of the  {\em initial} pre-existing asymmetry, $N_{B-L}^{\rm p,i}$,
between the end of inflation and the onset of leptogenesis.
This would translate, in the absence of any wash-out, into 
a sizeable value of $\eta_B^{\rm p}$ comparable or greater than $\eta_B^{\rm exp}$.
The strong thermal leptogenesis condition requires  that
this initial value of the pre-existing asymmetry is efficiently washed out by the RH
neutrinos wash-out processes in a way that the final value of $\eta_B$ is
dominated by $\eta_B^{\rm lep}$.\footnote{For definiteness we adopt a criterium $\eta_B^{\rm p} < 0.1\,\eta_B^{\rm lep}$.
In any case  the
constraints on low energy neutrino parameters depend only logarithmically on the  precise
maximum allowed value for $\eta_B^{\rm p}/\eta_B^{\rm lep}$.}
The predicted value of the baryon-to-photon number ratio is then 
dominated by the contribution from leptogenesis, that can be calculated as \cite{pedestrians}
\be
\eta_B^{\rm lep} =a_{\rm sph}\,{N_{B-L}^{\rm lep,f}\over N_{\g}^{\rm rec}} \simeq 
0.96\times 10^{-2}\,N_{B-L}^{\rm lep,f} \,  ,
\ee
accounting for sphaleron conversion \cite{sphalerons} and photon dilution
and where, in the last numerical expression, we normalised the abundance $N_X$ of some generic quantity $X$  
in a way that the ultra-relativistic equilibrium abundance of a RH neutrino 
$N_{N_i}^{\rm eq}(T \gg M_i) =1$. 
Successful leptogenesis requires that $\eta_B^{\rm lep}$ 
reproduces the experimental value in Eq.~(\ref{etaBexp}).

We can give analytical expressions for both two terms in Eq.~(\ref{2terms})
valid for a hierarchical RH neutrinos mass spectrum as implied by the Eqs.~(\ref{RHspectrum})
that leads to a $N_2$-dominated scenario of leptogenesis.\footnote{As discussed in detail in 
\cite{VLimpact}, a compact spectrum solution \cite{afs,compact}
with $M_1 \sim M_2 \sim M_3 \sim 10^{10-12}\,{\rm GeV}$ is also possible if, as it can be seen from Eqs.~(\ref{RHspectrum}),  ${\cal O}((\widetilde{m}_{\nu})_{11}) \ll {\cal O}(1$--$10\,{\rm meV})$  and $
{\cal O}(1/(\widetilde{m}_{\nu}^{-1})_{33}) \gg {\cal O}(1-10\,{\rm meV})$ while
$(\widetilde{m}_{\nu})_{11}/(\widetilde{m}_{\nu}^{-1})_{33}\sim~ (1$--$100) \, {\rm meV}^2$. 
In this case, however, it follows from Eq.~(\ref{Omegaapp})
and from the meaning of orthogonal matrix \cite{geometry}, that the seesaw formula
 implies huge fine-tuned cancellations to reproduce the measured 
 solar and atmospheric neutrino mass scales,
as discussed  in \cite{kingdibari}. However, recently a (string D-brane) model 
has been proposed in \cite{napolistring} where a compact spectrum 
emerges naturally. This is also  an example of a $SO(10)$-inspired model
that is not a $SO(10)$ model.}
The relic value of the pre-existing asymmetry 
is the sum of three
contributions from each flavour $N_{B-L}^{\rm p, f} = \sum_{\a} \,  N_{\D_\a}^{\rm p,f}$,  
whose expressions are given by
\bea\label{finalpas}
N_{\D_\t}^{\rm p,f} & = & 
(p^0_{{\rm p}\t}+\D p_{{\rm p}\tau})\,  e^{-{3\pi\over 8}\,(K_{1\t}+K_{2\t})} \, N_{B-L}^{\rm p,i} \,  , \\  \nonumber
N_{\D_\m}^{\rm p,f} & = & 
\left\{(1-p^0_{{\rm p}\t})\,\left[
p^0_{\mu\t_2^{\bot}}\, p^0_{{\rm p}\t^\bot_2}\,
e^{-{3\pi\over 8}\,(K_{2e}+K_{2\m})} + (1-p^0_{\m\t_2^{\bot}})\,(1-p^0_{{\rm p}\t^\bot_2}) \right]  + 
\D p_{{\rm p}\mu}\right\}
\,e^{-{3\pi\over 8}\,K_{1\m}}\, N_{B-L}^{\rm p,i}
 ,  \\  \nonumber
N_{\D_e}^{\rm p,f}& = & 
\left\{(1-p^0_{{\rm p}\t})\,\left[ 
p^0_{e\t_2^{\bot}}\,p^0_{{\rm p}\t^\bot_2}\,
e^{-{3\pi\over 8}\,(K_{2e}+K_{2\m})} + (1-p^0_{e \t_2^{\bot}})\,(1-p^0_{{\rm p}\t^\bot_2}) \right]  + \D p_{{\rm p} e}\right\}
 \,e^{-{3\pi\over 8}\,K_{1e}} \,\, N_{B-L}^{\rm p,i} \,   .
\eea
In this expression the $K_{i\a}$ are the {\em flavoured decay parameters} defined as
\be
K_{i\a} \equiv {\G_{i\a}+\overline{\G}_{i\a}\over H(T=M_i)}= 
{|m_{D\a i}|^2 \over M_i \, m_{\star}} \,  ,
\ee
where  $\Gamma_{i\a}=\Gamma (N_i \ra \phi^\dagger \, l_\alpha)$ 
and $\bar{\Gamma}_{i \a}=\Gamma (N_i \ra \phi \, \bar{l}_\alpha)$ are the
zero temperature limit of the flavoured decay rates into $\a$ leptons
and anti-leptons in the three-flavoured regime, $m_{\star}\simeq 1.1 \times 10^{-3}\,{\rm eV}$
is the equilibrium neutrino mass, $H(T)=\sqrt{g^{SM}_{\star}\,8\,\pi^3/90}\,T^2/M_{\rm P}$ is the expansion rate
and $g_{\star}^{SM}=106.75$ is the number of ultra-relativistic degrees of freedom in the standard model.
Using the bi-unitary parameterisation Eq.~(\ref{svd}) for $m_D$,  
the flavoured decay parameters can be expressed as 
\be\label{KialVL}
K_{i\a} = {\sum_{k,l} \, 
m_{Dk}\, m_{Dl} \,V_{L k\a} \, V_{L l \a}^{\star} \, U^{\star}_{R ki} \, U_{R l i} 
\over M_i \, m_{\star}}\,  .
\ee
In Eq.~(\ref{finalpas}) the quantities $p^0_{{\rm p} \tau}$ and $p^0_{{\rm p}\tau_2^{\bot}}$ 
are the fractions of the initial pre-existing asymmetry in the tauon flavour and 
in the flavour $\tau_2^{\bot}$, where $\tau_2^{\bot}$  is the  electron and muon flavour  superposition component in the leptons  
produced by $N_2$-decays (or equivalently the flavour component that is 
washed-out in the inverse processes producing $N_2$) so that $p^0_{{\rm p} \tau}+
p^0_{{\rm p}\tau_2^{\bot}} =1$.
The two quantities $p^0_{\alpha \t_2^{\bot}} \equiv K_{2\alpha}/(K_{2e}+K_{2\mu}) \; (\a=e,\mu)$ are
then the fractions of $\a$ pre-existing asymmetry in the  $\tau_2^{\bot}$ component, so that
$p^0_{e \t_2^{\bot}}+p^0_{\mu \t_2^{\bot}} =1$.

The contribution from leptogenesis also has to be calculated as the sum of
the three contributions from each flavour, explicitly \cite{bounds,vives,fuller,density}
\bea\label{twofl} \nonumber
N_{B-L}^{\rm lep, f} & \simeq &
\left[{K_{2e}\over K_{2\tau_2^{\bot}}}\,\ve_{2 \tau_2^{\bot}}\kappa(K_{2 \tau_2^{\bot}}) 
+ \left(\ve_{2e} - {K_{2e}\over K_{2\tau_2^{\bot}}}\, \ve_{2 \tau_2^{\bot}} \right)\,\kappa(K_{2 \tau_2^{\bot}}/2)\right]\,
\, e^{-{3\pi\over 8}\,K_{1 e}}+ \\ \nonumber
& + &\left[{K_{2\mu}\over K_{2 \tau_2^{\bot}}}\,
\ve_{2 \tau_2^{\bot}}\,\kappa(K_{2 \tau_2^{\bot}}) +
\left(\ve_{2\mu} - {K_{2\mu}\over K_{2\tau_2^{\bot}}}\, \ve_{2 \tau_2^{\bot}} \right)\,
\kappa(K_{2 \tau_2^{\bot}}/2) \right]
\, e^{-{3\pi\over 8}\,K_{1 \mu}}+ \\
& + &\ve_{2 \tau}\,\kappa(K_{2 \tau})\,e^{-{3\pi\over 8}\,K_{1 \tau}} \,  ,
\eea
where  $\ve_{2\a} \equiv -(\G_{2\a}-\overline{\G}_{2\a})/(\G_2 + \overline{\G}_2)$ are the $N_2$ flavoured $C\!P$ asymmetries, with $\G_2 \equiv \sum_{\a} \G_{2\a}$ and
$\overline{\G}_{2}\equiv \sum_{\a} \, \overline{\G}_{2\a}$.
Using the singular value decomposition Eq.~(\ref{svd}) for $m_D$, 
the flavoured $C\!P$ asymmetries can be calculated using the approximate expression \cite{VLimpact}
 \be\label{ve2al}
\ve_{2\a} \simeq {3 \over 16\, \pi \, v^2}\,
{|(\widetilde{m}_{\nu})_{11}| \over m_1 \, m_2 \, m_3}\,
{\sum_{k,l} \,m_{D k} \, m_{Dl}  \,  {\rm Im}[V_{L k \a }  \,  V^{\star}_{L  l \a } \, 
U^{\star}_{R k 2}\, U_{R l 3} \,U^{\star}_{R 3 2}\,U_{R 3 3}] 
\over |(\widetilde{m}_{\nu}^{-1})_{33}|^{2} + |(\widetilde{m}_{\nu}^{-1})_{23}|^{2}}   \,  .
\ee
In the case of strong thermal leptogenesis, the final asymmetry has to be necessarily\footnote{If $\ell_2$ coincides with great precision with the electron or muon flavour,
then it is possible in principle to have very special electron or muon dominated solutions.}
tauon dominated \cite{problem} and the previous expression reduces simply to 
\be\label{taufinal}
N_{B-L}^{\rm lep, f} \simeq \ve_{2\t} \, \k (K_{2\t}) \, e^{-{3\pi \over 8}\, K_{1\t}} \,   .
\ee

\section{The  benchmark case: $N^{\rm p, i}_{B-L} = 10^{-3}$ and $\a_2 =5$}

The set of analytical expressions given in the previous sections
allows an efficient analytic calculation of the asymmetry that avoids the lengthy numerical diagonalisation of the 
Majorana mass matrix in the Yukawa basis (see Eq.~(\ref{invM})). We have run a Montecarlo to 
derive the allowed region in the space of parameters for the benchmark case $N^{\rm p, i}_{B-L} = 10^{-3}$ and $\a_2 =5$.
The results, shown in Fig.~1, are projected on  different planes: in the top panel in the plane $\d$ versus $\theta_{23}$,
in the central panel in the plane $\rho$ versus $\sigma$, in the 
bottom panel in the plane $m_{ee}$ versus $m_1$.
\begin{figure}
\begin{center}
\psfig{file=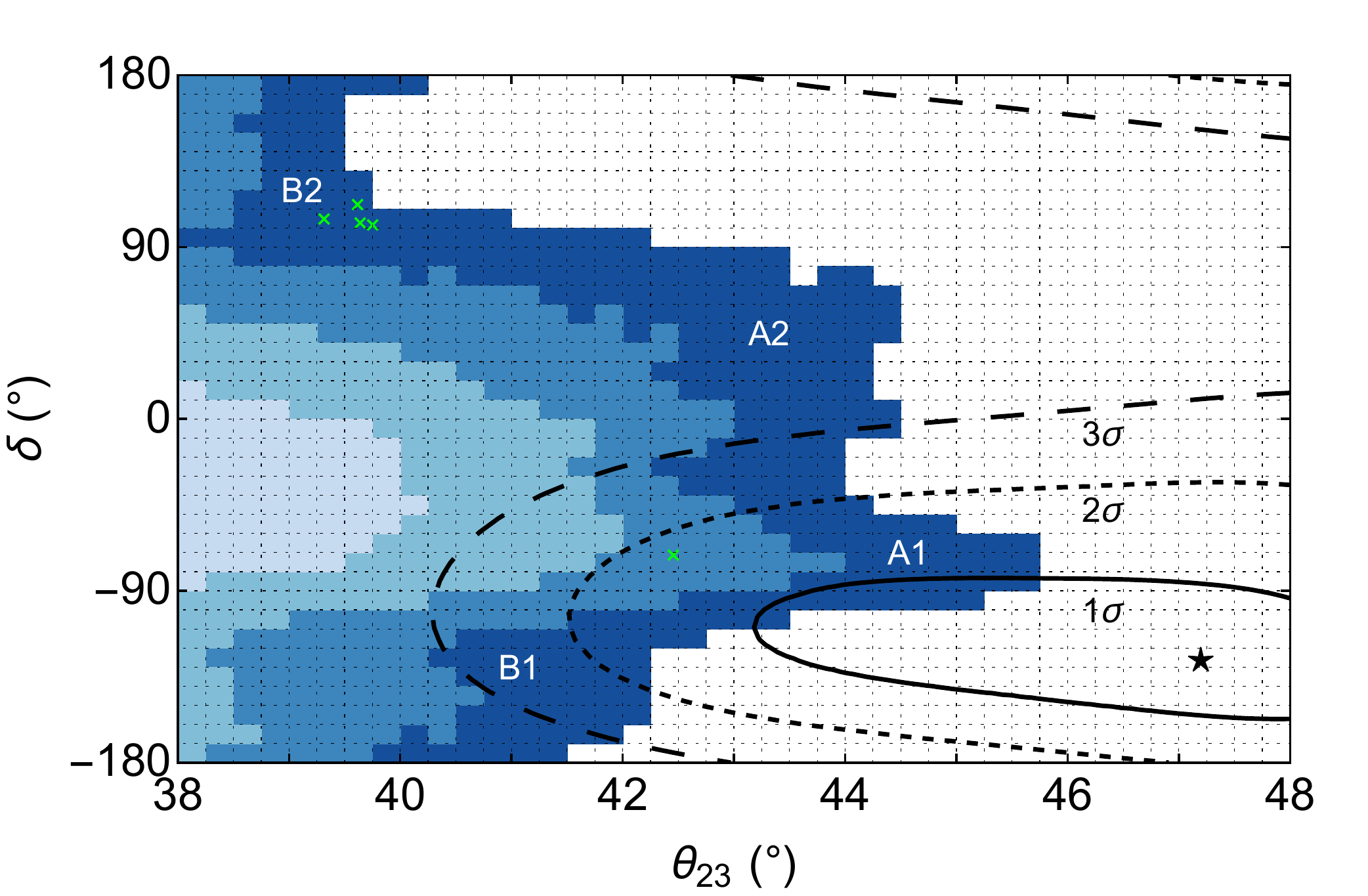,height=55mm,width=85mm} \\
\vspace{-3mm}
\psfig{file=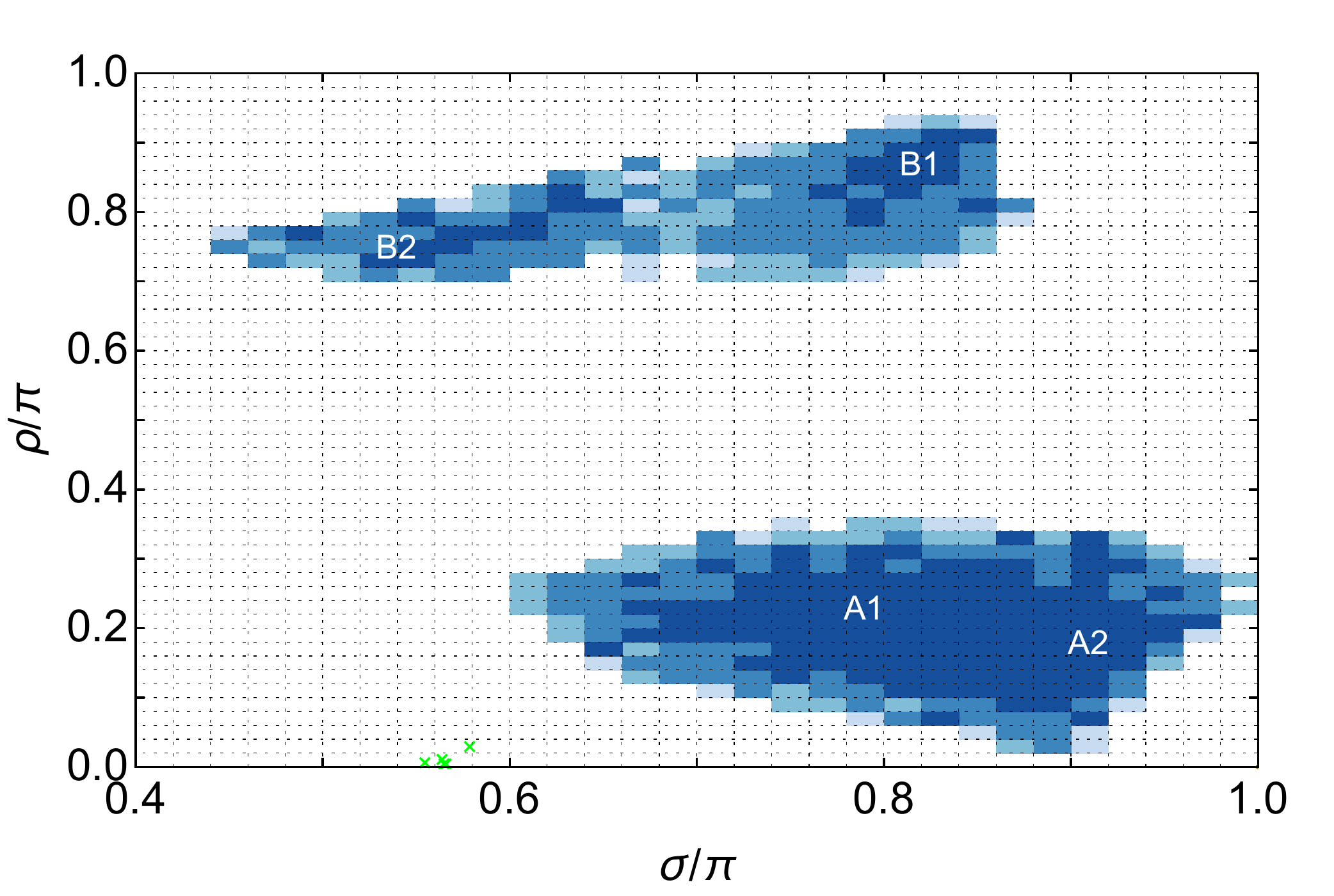,height=55mm,width=85mm}  \\
\vspace{-3mm}
\psfig{file=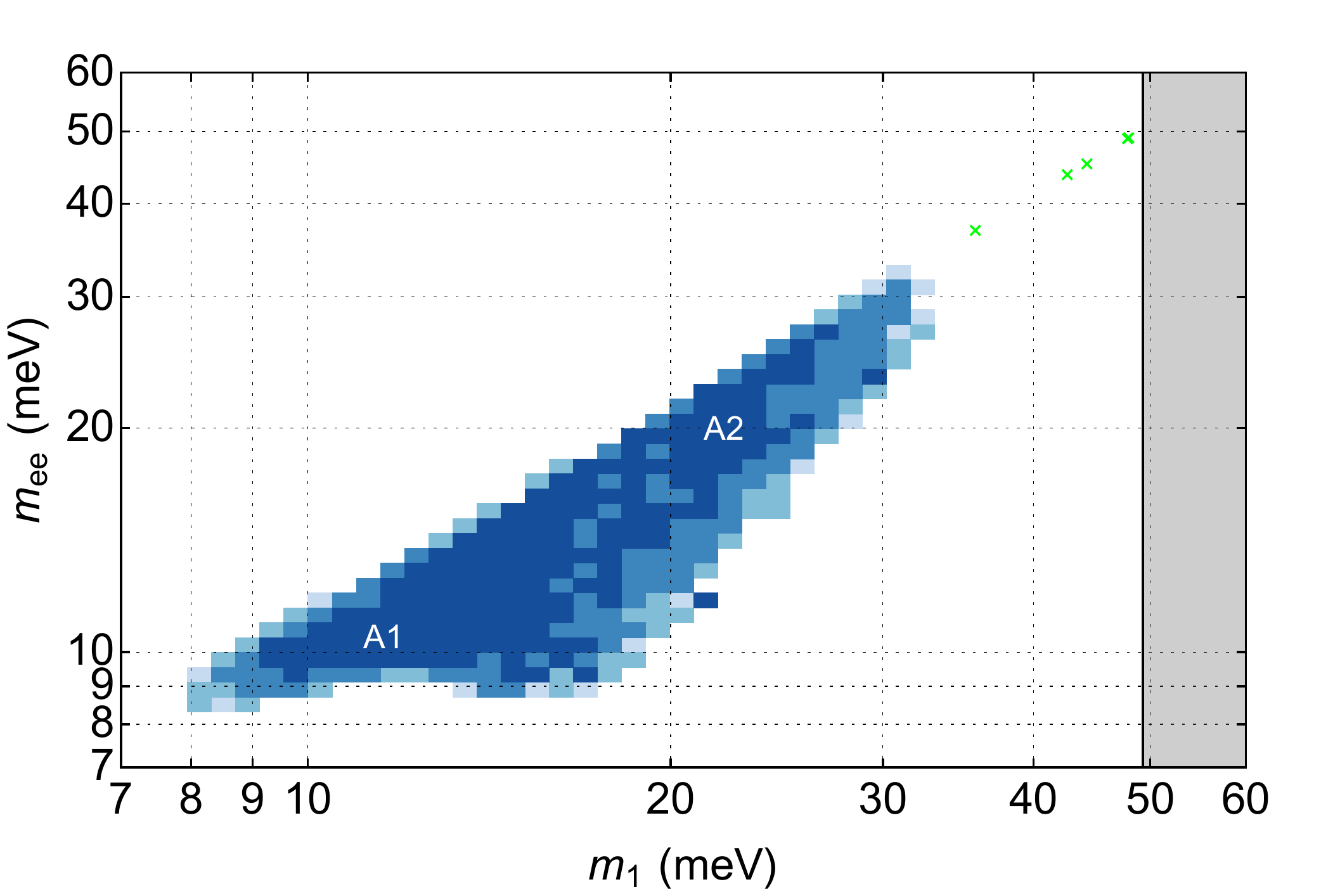,height=55mm,width=85mm}   
\end{center}
\vspace{-8mm}
\caption{STSO10 leptogenesis allowed region projected on three different low energy neutrino
parameters planes for the benchmark case $\a_2=5$ and $N_{B-L}^{\rm p,i}=10^{-3}$. 
The  regions with different blue graduation  indicate the 
regions containing $68\%$, $95\%$, $99.7\%$ and $100\%$ of the solutions (respectively from lightest to darkest blue) in the plane $\delta$ versus $\theta_{23}$. 
The green crosses correspond to special muon-dominated solutions. In the top panel, the black lines represent the results of the latest global neutrino analysis \cite{nufit} with the best fit indicated by the star. 
The gray band in $m_1$ is the excluded  region from the cosmological upper bound on $m_1$ from {\em Planck}.}
\label{benchmark}
\end{figure}
The  plots in the figure have been obtained finding ${\cal O}(10^6)$ solutions out of ${\cal O}(10^{11})$ runs
where unknown (or poorly known) parameters $\log_{10}[m_1]$, $\rho$, $\sigma$, $\theta_{23}$, $\d$ and the 6 parameters in $V_L$ 
have been generated uniformly randomly except for the measured parameters
$m_{\rm atm}$, $m_{\rm sol}$, $\theta_{12}$, $\theta_{13}$ that have been generated Gaussian randomly
around their best fits. 
The results confirm the gross features found in previous papers \cite{strongSO10,SO10decription,VLimpact} but
the much higher number of solutions (about 3 orders of magnitude) we found has made possible to 
saturate\footnote{This has been done by first running uniformly on all parameter space until the probabilities in each bin 
with $\Delta \d = 10^{\circ}$ and $\Delta \theta_{23} =0.25^{\circ}$ became stable except in bins at the contour
and then running additional Montecarlo's focussing on  bins at the contour.   If a new solution was found the procedure 
was repeated until saturation. We have been extremely carefully in determining the bound in the region experimentally 
allowed at large $\theta_{23} \gtrsim 42^{\circ}$ and $\delta \lesssim -30^{\circ}$. }
the allowed regions, for example determining the upper bound on $\theta_{23}$ with much higher accuracy. 
Just for the benchmark case we have double checked the constraints obtained by using the analytical expression and the 
calculation of the asymmetry through a numerical diagonalisation of $M^{-1}$.
This further confirms the validity of the general analytical solution found in \cite{SO10decription,VLimpact}.

We have also linked the projections of the solutions  in the plane $\delta$ versus $\theta_{23}$
with those in the plane $\rho$ versus $\sigma$. This shows that the two 
disjoint allowed regions in the plane $\rho$ versus $\sigma$, a dominant one at low  values of $\rho$  
($\rho\simeq [0.05\pi, 0.30\pi] +n\,\pi$ with $n$ integer) 
and a sub-dominant one at high values of $\rho$ ($\rho\simeq [0.70\pi, 0.90\pi] +n\,\pi$ with $n$ integer), correspond
in the plane $\delta$ versus $\theta_{23}$  to partially overlapping regions  upper bounded respectively by $\theta_{23} \lesssim 45.75^{\circ}$, 
with the upper bound saturated at $\delta \simeq -75^{\circ}$,
and by $\theta_{23} \lesssim 42^{\circ}$ with the upper bound
saturated at $\delta \simeq-120^{\circ}$.  These two sets of solutions  are indicated in the panels
respectively as region $A$ and region $B$.  The two regions can be further decomposed into two subregions 
for low and high values of $\sigma$, that we indicate in the figure respectively with $A_1$, $A_2$ and $B_1$, $B_2$,
where $A_1$ dominates over $A_2$ and $B_1$ over $B_2$. The region $A_2$ is upper bounded by $\theta_{23} \lesssim 44^{\circ}$
and the upper bound is saturated at $\delta \simeq 60^{\circ}$. This region is now ruled out at $99\%$C.L. by the new experimental 
constraints. The region $B_2$ is ruled out at much more than $99\%$C.L.. 
The region $B_1$ is also ruled out at $95\%$C.L. Hence, with the new results basically
only the region $A_1$ is allowed for $\delta =[ -120^{\circ}, -30^{\circ}]$. 

In the panels of Fig.~1  we have also indicated with 4 different blue graduations 
the regions containing $68\%$, $95\%$, $99.7\%$ and $100\%$ of the solutions (respectively from lightest to darkest blue) in the plane $\delta$ versus $\theta_{23}$. 
These regions of course depend on how the parameters are randomly generated, in our case uniformly in the mixing angles and phases.\footnote{One could have also used different ways, for example uniformly in $\sin\theta_{ij}$ or $\sin^2\theta_{ij}$.}
However, they provide a good indication of when solutions start to get fine-tuned especially in the values of the Majorana phases as one can see from the central panel. In particular, the asymmetry is suppressed approximately as
$\propto \sin \theta_{23}^{-4}$~\cite{SO10decription}
and going at larger values of $\theta_{23}$ all allowed ranges of parameters shrink around the values
that maximise the asymmetry up to a maximum value of $\theta_{23}$ that determines an upper bound
(of course depending on $\delta$). Therefore, increasing values of $\theta_{23}$ implies
a higher and higher fine-tuning  of all parameters to realise STSO10.  
As an example in Fig.~2 we plot the dependence of $\eta_B^{\rm lep}$
on $m_1$ for four different values of the angle $\theta_{23}$ (the color code of the four lines in the plot refers to the different regions in the panel of Fig.~1), while the values of all other parameters are fixed (see the figure caption). In particular, $\theta_{23} \simeq 45.75^{\circ}$ is the highest value we found in correspondence of $\delta \simeq - 75^{\circ}$. This plot provides a good idea of the amount of fine-tuning implied by these marginal solutions.
\begin{figure}
\begin{center}
\psfig{file=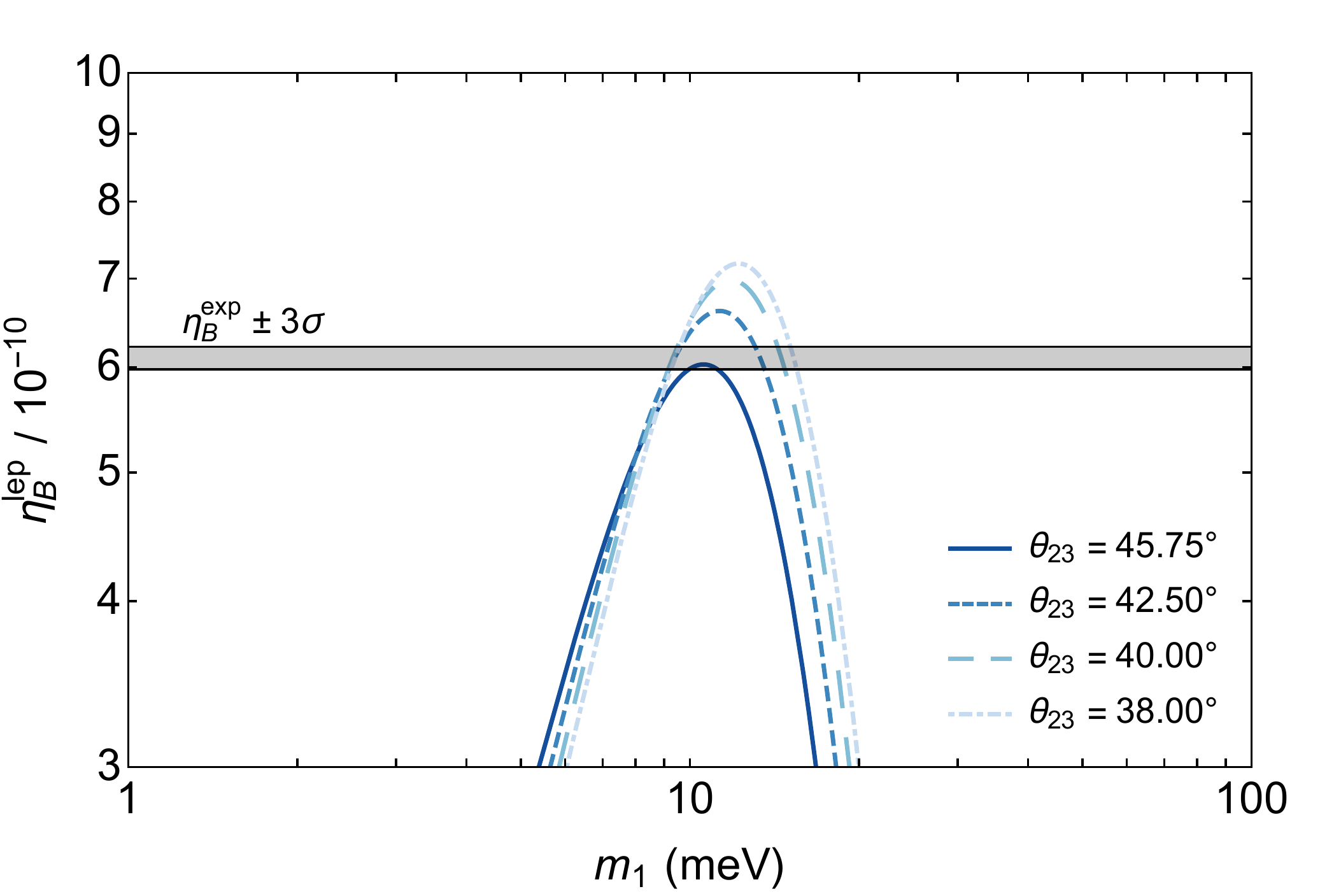,height=55mm,width=85mm} 
\end{center}
\vspace{-8mm}
\caption{Plot of the final asymmetry from leptogenesis as a function of the neutrino mass $m_1$. The different lines correspond to the four indicated values of  $\theta_{23}$ each for one of the four regions in Fig.~1. The other neutrino parameters are fixed to: $\delta \simeq -75^{\circ}$, $\rho \simeq 0.23\,\pi$, $\sigma \simeq 0.84\,\pi$, $\rho_L\simeq 0.06\,\pi$, $\sigma_L \simeq 1.1\,\pi$, $\delta_L \simeq -0.47 \,\pi$, $\theta_{13}^L\simeq 0.08^{\circ}$, $\theta_{23}^L \simeq 2.2 ^{\circ}$ and $\theta_{12}^L \simeq 12.1^{\circ}$.}
\label{DifferentPre}
\end{figure}
Analogously, a higher and higher fine-tuning is also required for $\delta$ more and more outside the bulk region falling mainly in the 4th quadrant solutions, especially at large values $\theta_{23} \gtrsim 42^{\circ}$, now favoured by most recent experimental results.

Before concluding this section we just observe that we found few special muon-dominated solutions living at very large values of $m_1$. Hence, this muonic solution are only marginally allowed by the current cosmological upper bound.\footnote{The existence of such special muonic solutions had been found in \cite{strongSO10}.
They are accidental and correspond to the case when the flavour ${\ell}_2$ is very precisely aligned along the muon flavour.}
These are indicated by the green crosses  in all panels of Fig.~1. 

\section{Dependence of the constraints on $N^{\rm p,i}_{B-L} $ and $\a_2$}

We have studied the dependence of the constraints on the two parameters $N_{B_L}^{\rm p,i}$ and $\a_2$, 
the first related to the history of the very early universe prior to leptogenesis, the second to neutrino properties. 

We have first studied the variation with $N_{B-L}^{\rm p,i}$  for $\a_2=5$. The results are shown 
in Fig.~3. One can see how the allowed regions shrink in all planes for increasing value of $N_{B-L}^{\rm p,i}$. In particular, the case $N_{B-L}^{\rm p,i} =10^{-2}$ survives at $95\% {\rm C.L.}$ only for a very marginal region while the case $N_{B-L}^{\rm p,i} = 0.1$ survives marginally only at $99\%\,{\rm C.L.}$
\begin{figure}
\begin{center}
\psfig{file=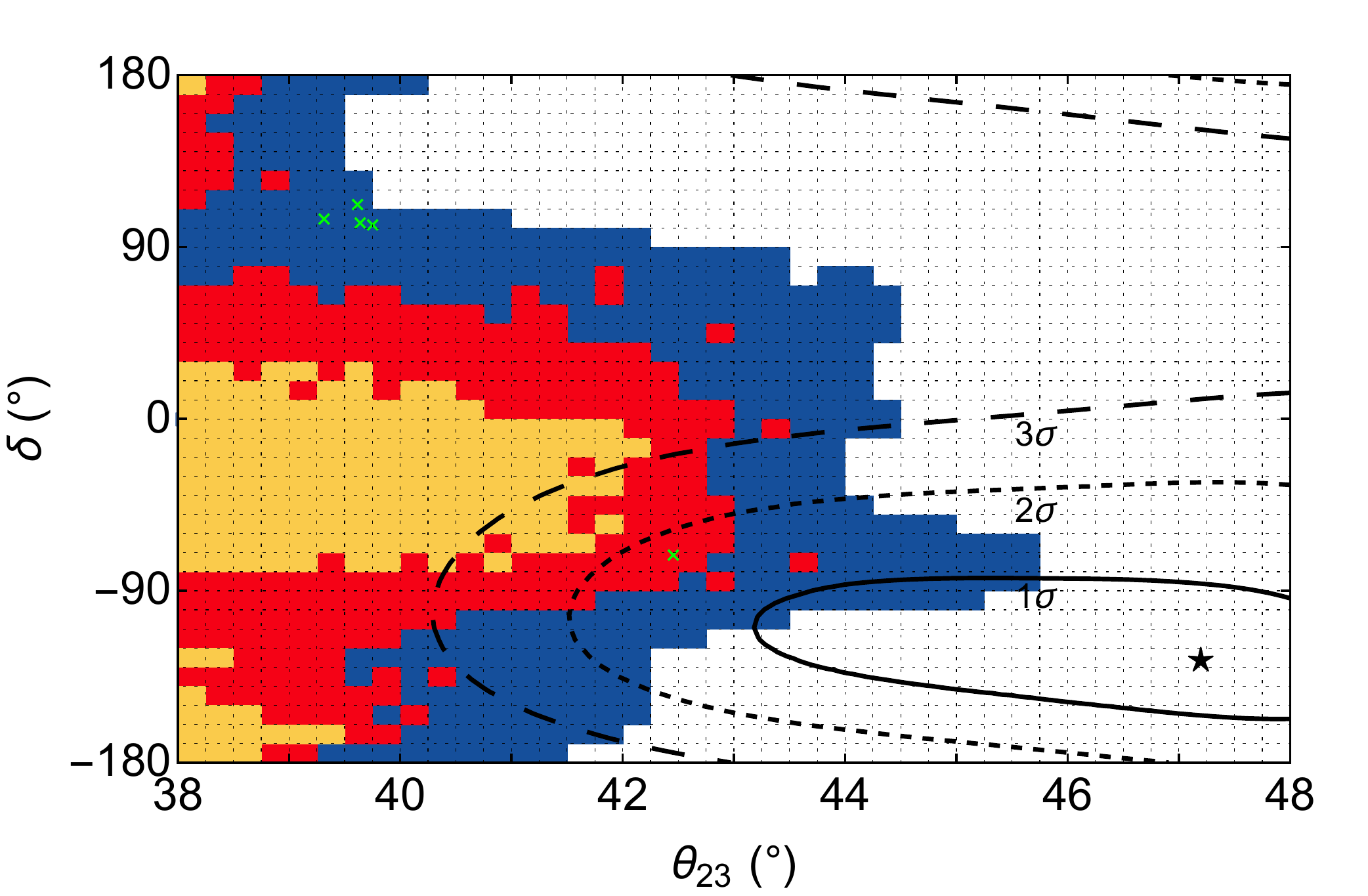,height=55mm,width=85mm} \\
\psfig{file=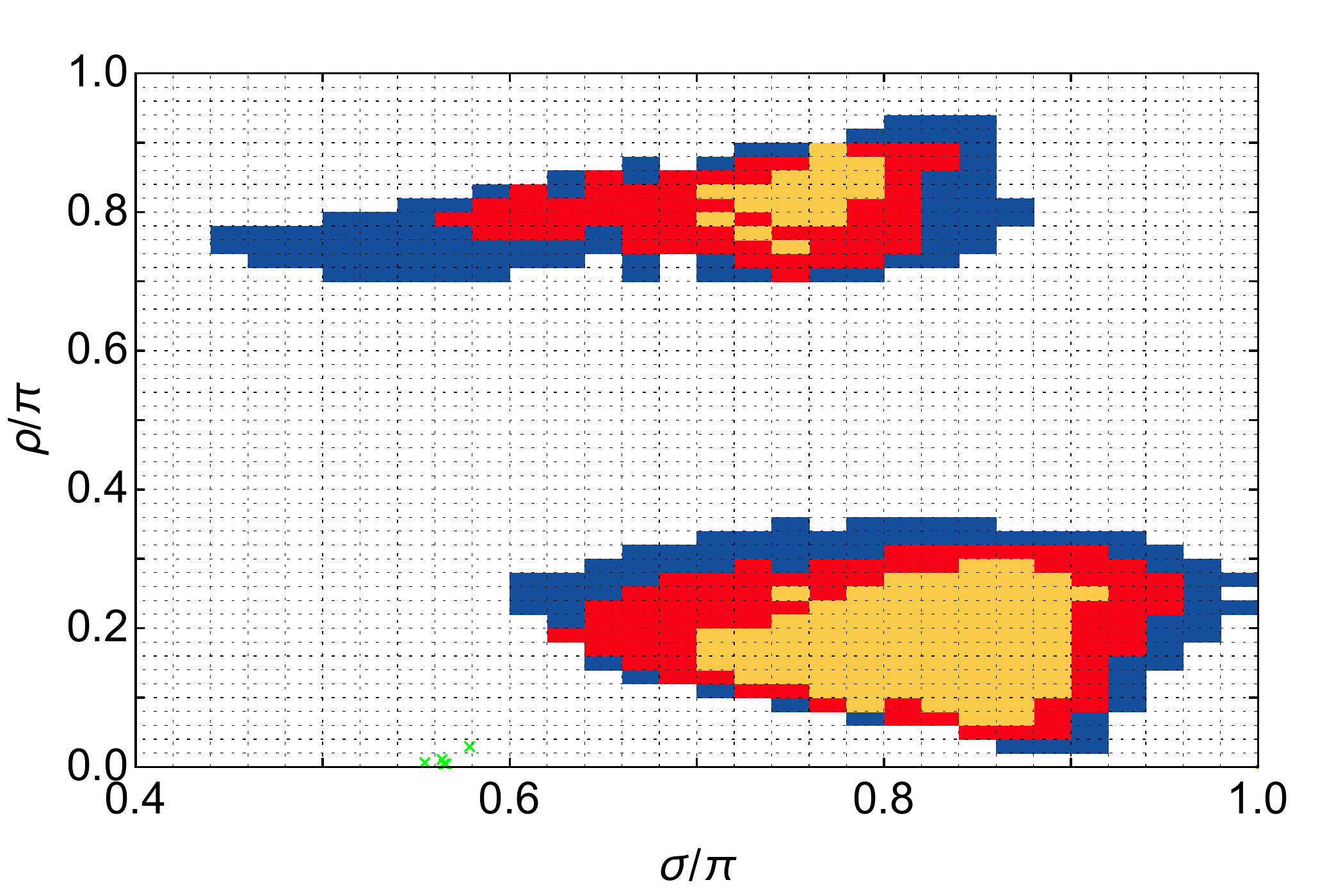,height=55mm,width=85mm} \\
\psfig{file=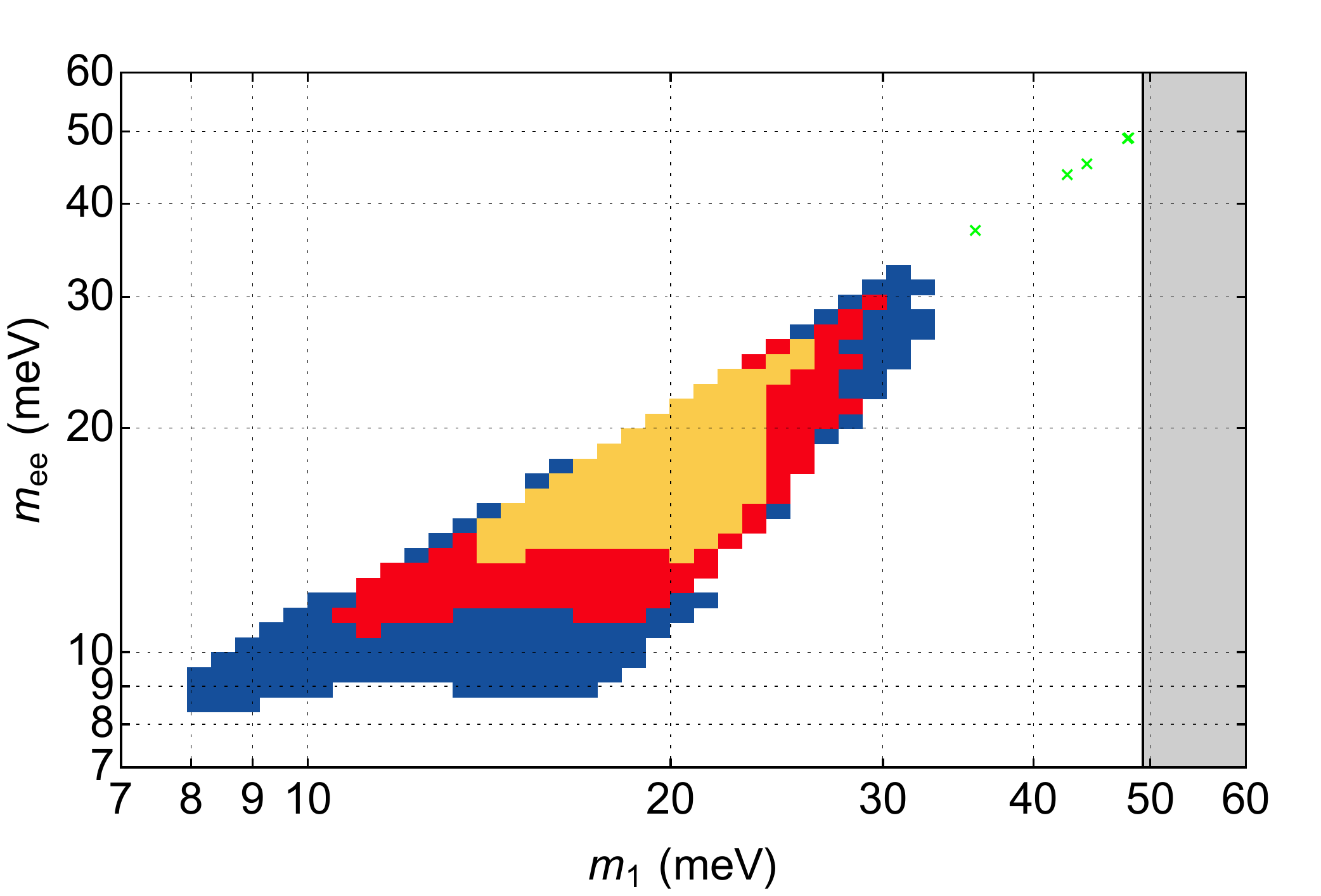,height=55mm,width=85mm} 
\end{center}
\vspace{-8mm}
\caption{Dependence of the allowed region on $N_{B-L}^{\rm p,i}$ for $\a_2=5$ and three
different values of $N_{B-L}^{\rm p,i}$: $10^{-1}$ (yellow), $10^{-2}$ (red), $10^{-3}$ (blue).}
\label{etaBth23max}
\end{figure}
\begin{figure}
\begin{center}
\psfig{file=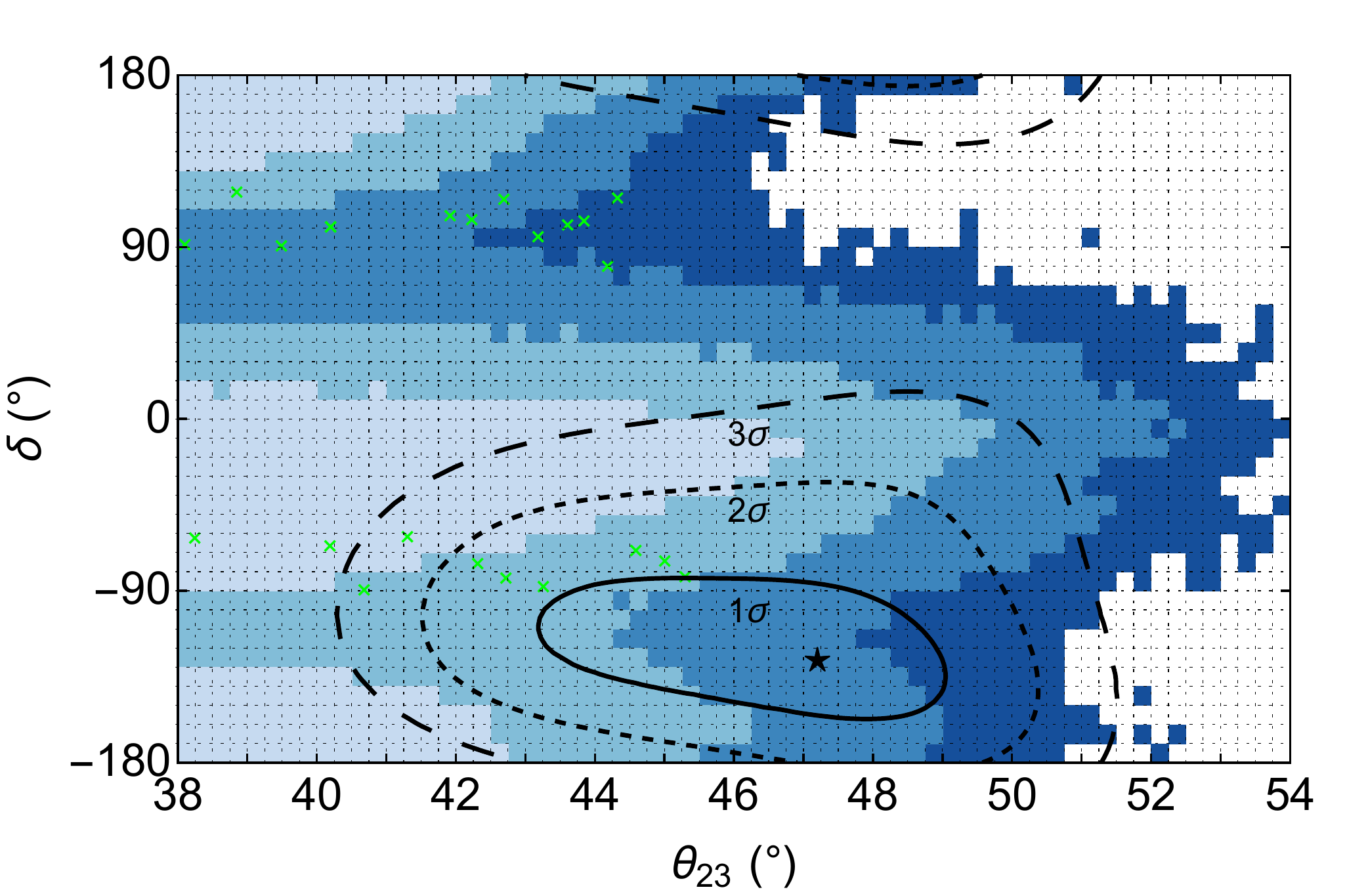,height=55mm,width=85mm} \\
\psfig{file=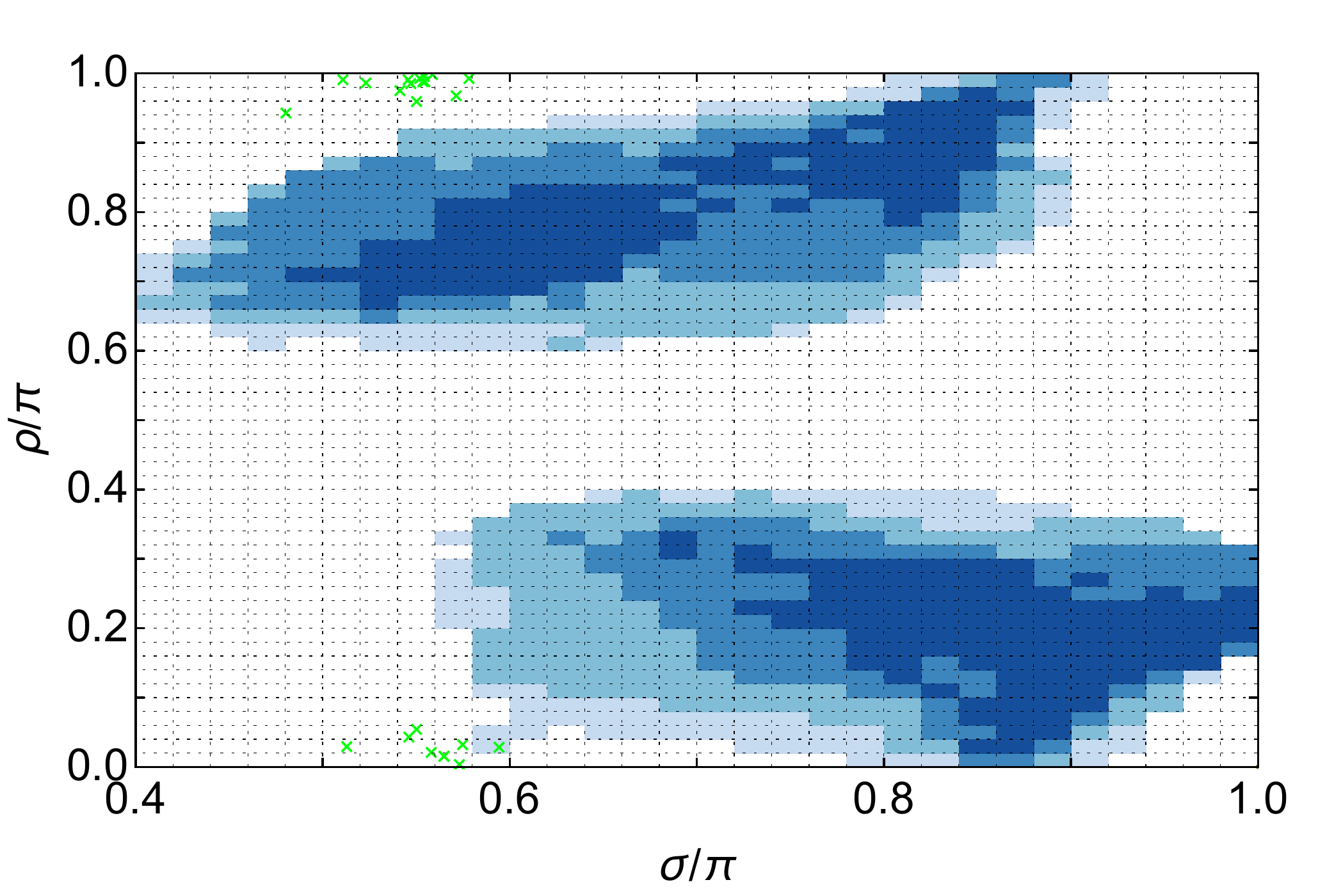,height=55mm,width=85mm} \\
\psfig{file=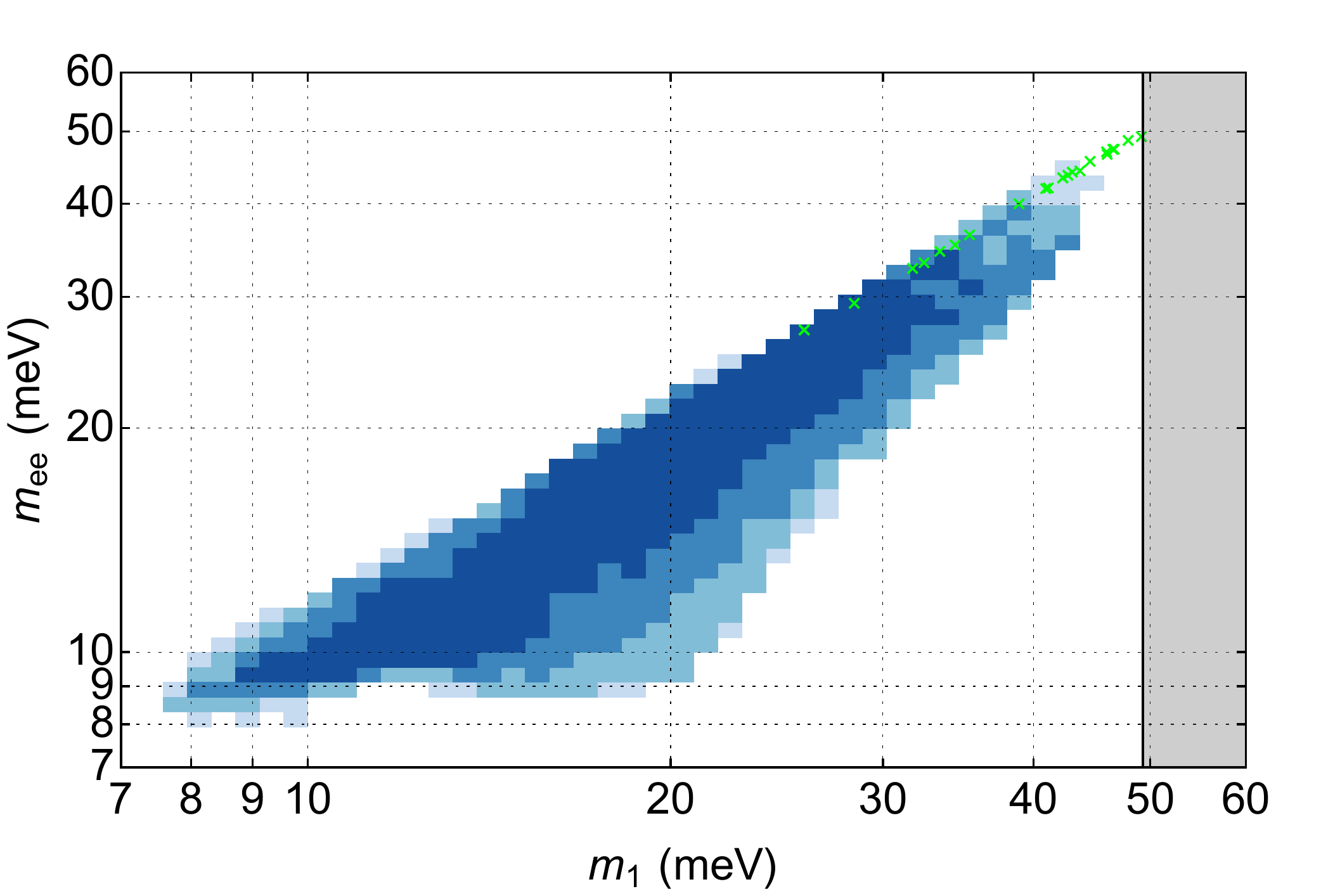,height=55mm,width=85mm} 
\end{center}
\vspace{-8mm}
\caption{Allowed region for $\a_2=6$ and $N_{B-L}^{\rm p,i}=10^{-3}$, with same colour code as in Fig.~1.}
\label{fig:al6}
\end{figure}
Then, in Fig.~4, we provide the results for $\a_2 =6$ and $N_{B-L}^{\rm p,i}=10^{-3}$ with the same colour code of Fig.~1.
As it can be seen the allowed region shifts toward higher values of $\theta_{23}$.
When one considers the bulk of the solutions (in lightest blue),  one can see that the allowed range
of $\delta$ at $95\%\,{\rm C.L.}$ is  $\delta \simeq [-70^{\circ},-30^{\circ}]$ and $\theta_{23} \lesssim 
46^{\circ}$.  
If one considers slightly more marginal solutions (next-to-lightest blue)  then also solutions in the third
quadrant for $\delta$ are found and $\theta_{23} \lesssim 50^{\circ}$.
Therefore, 
even if $\theta_{23}$ will be found in the second octant, STSO10 leptogenesis can work if $\a_2$ 
is large enough. The important point is then
how large $\a_2$ can be realistically. Here we notice that the value of $\a_2$ obtained in realistic fits seems 
to allow values $\a_2 \gg 1$. In a recent analysis \cite{rodejohann} fits have indeed been obtained with $a_2 \simeq 6$
and $\a_2 \simeq 8$.\footnote{Other recent realistic fits have been recently presented in \cite{babu}, but in this case 
$\a_2 \simeq 1.5$,  and in \cite{perdomo} where interestingly $\a_2 \simeq 6$ but in this case this value is indeed determined by successful leptogenesis condition.}
This possibility requires further investigation.

It is  interesting to notice that the lower bound on $m_{ee}$ does not depend on $\a_2$
and ultimately neutrinoless double beta decay  provides a crucial test for STSO10 since 
the range of allowed values is quite narrow, $m_{ee} \simeq\left[10,30\right]\,{\rm meV}$,  independently of $\a_2$.
Finally, in Fig.~5 we have summarised the results reporting the dependence of the upper bound on $\theta_{23}$ both on $\a_2$  and $N_{B-L}^{\rm p,i}$. 
In the left panel we have plotted the upper bound on $\theta_{23}$ for $\a_2=4,5,6$
as a function of $N_{B-L}^{\rm p,i}$.  One can see how the upper bound relaxes for
smaller values of $N_{B-L}^{\rm p,i}$. The grey band
indicates the experimental lower bound ($95\%\,{\rm C.L.}$).
 In the right panel we directly show the constraints in 
the plane $\a_2$ versus $N_{B-L}^{\rm p,i}$, indicating the region excluded at $95\%\,{\rm C.L.}$.
We can see that the current experimental constraints rule out a region 
$\a_2 \lesssim 4.7$ and  $N^{\rm p,i}_{B-L} \gtrsim 0.1$.
\begin{figure}
\begin{center}
\psfig{file=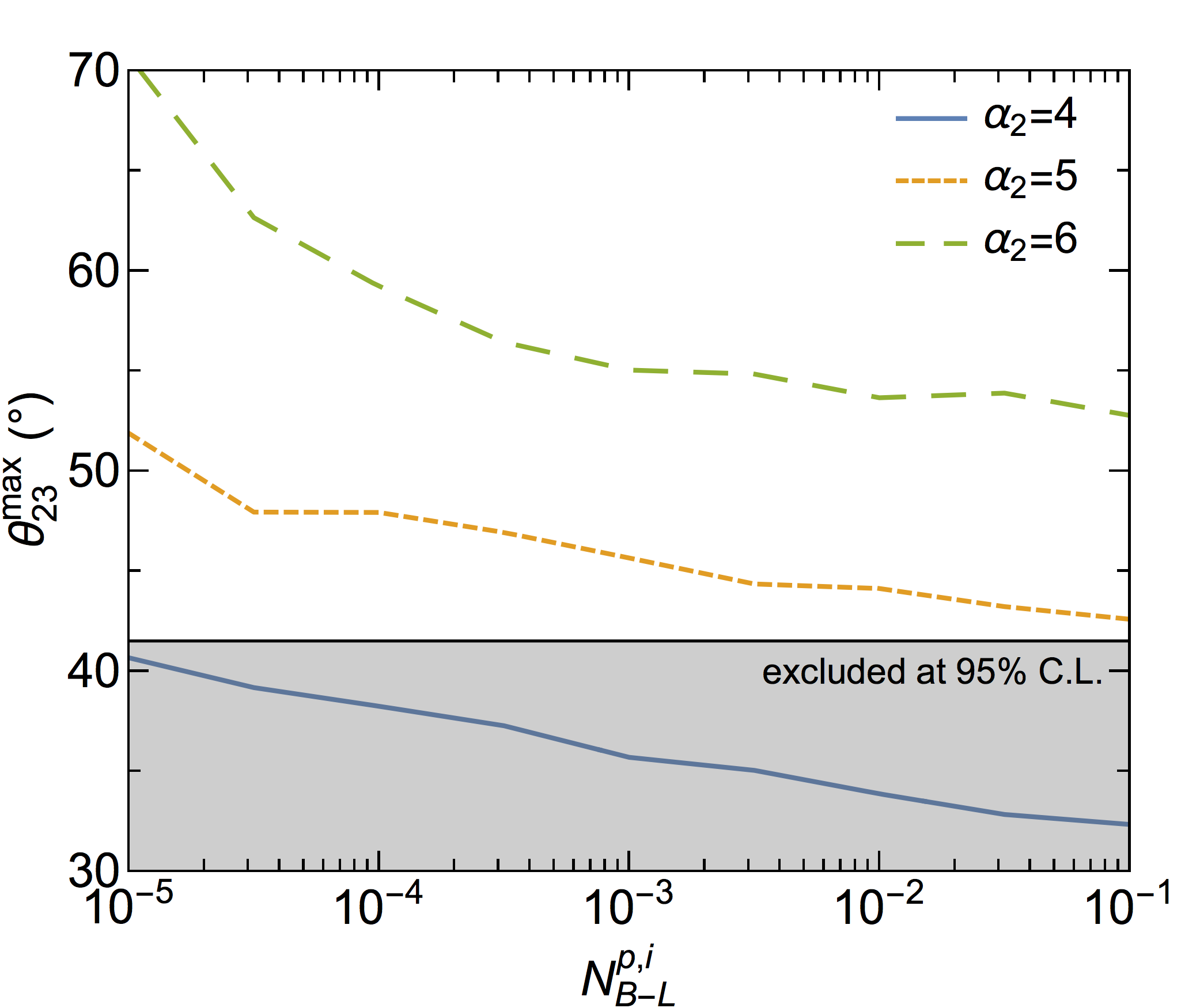,height=55mm,width=73mm} 
\psfig{file=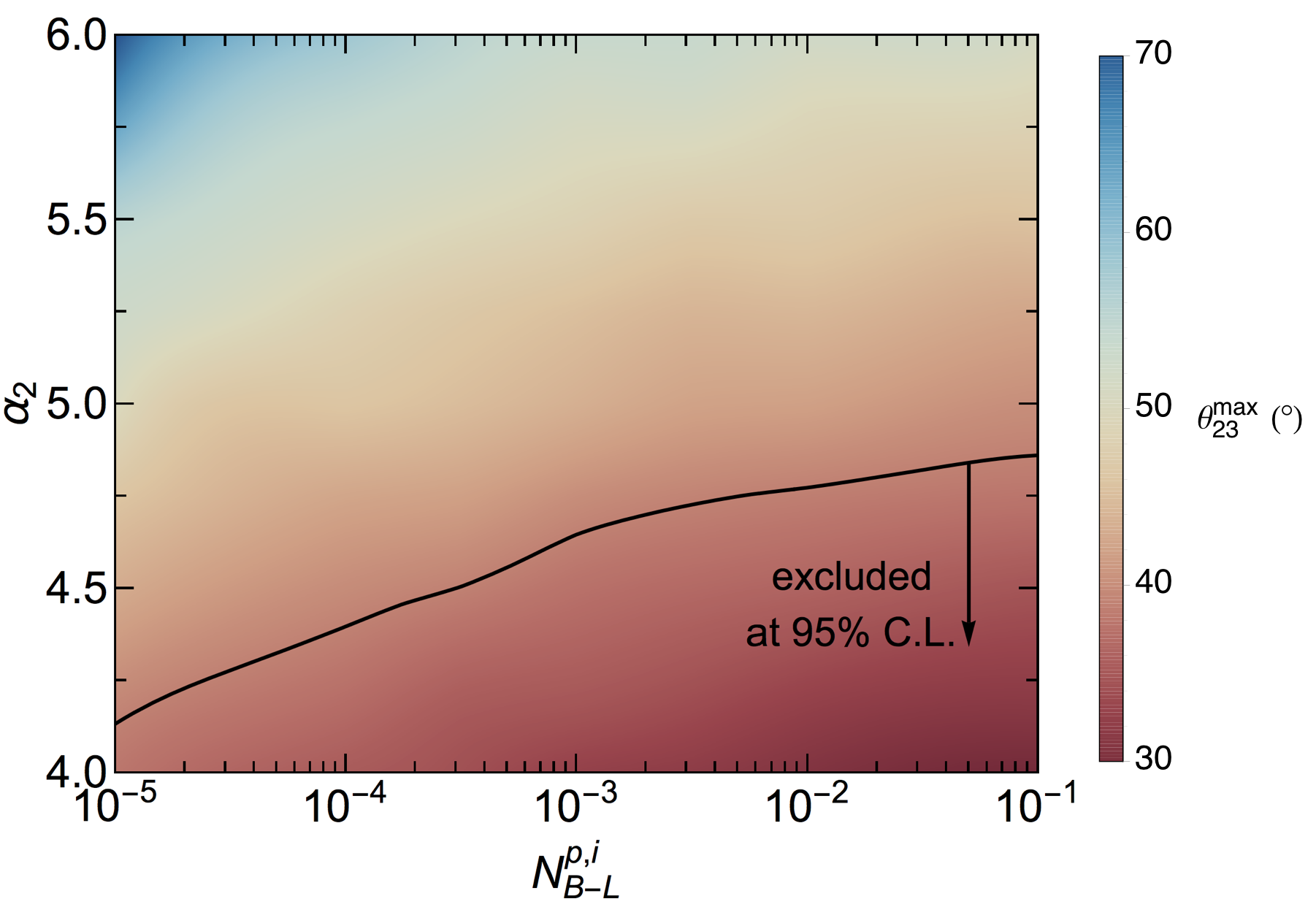,height=55mm,width=85mm}
\end{center}
\vspace{-8mm}
\caption{Left panel: Upper bound on $\theta_{23}$ as a function of $N_{B-L}^{\rm p,i}$, for three values of the parameter $\a_2$. The gray band shows the current $2\s$
experimental constraints $\theta_{23} \gtrsim 41.5^{\circ}$ \cite{nufit}.
Right panel: $95\% \,$C.L. constraint in the plane $\a_2$ versus $N_{B_L}^{\rm p,i}$.}
\label{theta23max}
\end{figure}

It is useful to give some analytical insight on the numerical results mainly based on the analyses presented
in \cite{SO10decription,VLimpact}. 
There are two conditions to be imposed: strong thermal leptogenesis and successful leptogenesis.
The first condition translates straightforwardly into a lower bound on $m_{ee}$, considering 
that $K_{1e} \simeq m_{ee}/m_{\star}$\footnote{This is strictly true in the approximation $V_L = I$ but it remains approximately valid also for
$V_L \lesssim V_{CKM}$.} 
and also that one has to impose $K_{1e} \gtrsim 8 + 0.85\,\ln(N_{B-L}^{\rm p,i}/10^{-3})$
translating into $m_{ee} \gtrsim 9\,{\rm meV}\,\left(1 + \ln(N_{B-L}^{\rm p,i}/10^{-3})\right)$.
This lower bound is well visible in the lowest panel of Fig.~3 where one can see how it is independent of $m_1$
and it increases logarithmically with $N_{B-L}^{\rm p,i}$. 
Notice that the $K_{i\a}$'s do not depend on the $\a_i$'s and in particular on $\a_2$, as 
it can be inferred easily from Eq.~(\ref{KialVL}). This implies that the strong thermal 
condition places constraints independent of $\a_2$.

In STSO10 leptogenesis the final asymmetry is dominated by the tauonic contribution 
 (see Eq.~(\ref{taufinal})). This is the product of three quantities: the $C\!P$ asymmetry $\ve_{2\t}$,
the efficiency factor $\k(K_{2\t})$ at the production and the wash-out factor from the lightest RH neutrino
$e^{-{3\pi \over 8}\,K_{1\t}}$. We can say approximately that the suppression of the asymmetry on $\sin\theta_{23}$
is mainly contained in $\ve_{2\tau}$ that in the approximation $V_L \simeq  I$ and for $m_1/m_{\rm sol} \ll 1$ 
is suppressed as $\sin\theta_{23}^{-4}$. In the case of STSO10 one has $m_1 \simeq m_{\rm sol}$ and the suppression
is milder but still present.
The dependence on the two Majorana phases and on the Dirac phase $\delta$ comes from an interplay between maximising
$\ve_{2\tau}$ and having $K_{1\tau} \lesssim 1$. The existence of two very well defined sets of solutions for different
values of $\rho$, the  ``A" region for $\rho \simeq 0.2\pi + n\,\pi$ and the  ``B" region for $\rho \simeq 0.8\,\pi + n\, \pi$,
is also quite well understood since in the limit $m_1/m_{\rm sol} \ll 1$ one has $\rho \rightarrow n\,\pi/2$ while for $m_1/m_{\rm sol} \gg 1$ one has $\rho \rightarrow n\,\pi$. For $m_1 \simeq m_{\rm sol}$, two solutions at intermediate value of $\rho$ are obtained below and above $0.5\,n\,\pi$, respectively. At the same time, approximately, in order to minimise $K_{1\t}$ and maximise $m_{ee}$,
one has respectively $2\,\s -\d \simeq 0$ and $\s-\d \simeq 0$ for $m_1/m_{\rm sol} \ll 1$. For $m_1 \simeq m_{\rm sol}$
again these two conditions split into two solutions, one for $2\s-\d <0$ corresponding to $\rho \simeq 0.2\,\pi + n\,\pi$ and $\d <0$
and one for $2\s-\d >0$ corresponding to $\rho \simeq 0.8\,\pi +n\,\pi$ and $\d >0$. The first solution is the dominant one
since it allows to maximise the asymmetry for $K_{1\t} \lesssim 1$. This translates into a dominance of the solution with $\d <0$. 
It should be noticed that also the conditions on the phases do not depend on $\a_2$ and this explains why the constraints on the phases
do not change with $\a_2$. The dependence on $\a_2$ can be entirely explained from the dependence 
$\ve_{2\t} \propto \a_2^2$ that translates into $\eta_B^{\rm lep} \propto \a_2^2$. Therefore, a higher value of $\a_2$ mainly relaxes
the upper bound on $\theta_{23}$ but not for example the lower bound on $m_{ee}$. This is why the range $m_{ee}\simeq [10,30]\,{\rm meV}$
can be regarded as quite a robust feature of STSO10 if one assumes $N_{B-L}^{\rm p,i} \gtrsim 10^{-3}$. This is interesting since
even for NO one expects a signal in future neutrinoless double beta decay experiments.

Before concluding this section it is useful to remind that there are some sources of theoretical uncertainties that might be relevant in the light of the fact that, as we have discussed, current experimental data seem to corner STSO10. First of all, we are neglecting the running of parameters and this might be particularly important for the value of $\a_2$ that is affected in particular by the running of $m_{\rm charm}$. We used an approximated value from \cite{charm} at a fiducial scale $\sim 10^{10}\,{\rm GeV}$ but, since constraints are particularly sensitive to $\alpha_2$, a more accurate determination of  $m_{\rm charm}$ at the precise scale of leptogenesis production for each solution might give some important effect.
Of course the running of neutrino parameters also should be taken into account, especially of the Dirac phase, since our constraints originate either
at the asymmetry production scale or at the lightest RH neutrino wash-out scale. Other effects we are neglecting
are flavour coupling \cite{fuller} and a more precise calculation of the asymmetry within a density matrix approach \cite{density}.

\section{Conclusion}

In this paper, we  presented a precise determination of the constraints on neutrino parameters from STSO10 leptogenesis, comparing them with recent experimental results from long baseline neutrino experiments. It is certainly encouraging that NO is favoured over IO at $\sim 2\,\s$ since this is a strict requirement for STSO10 leptogenesis. On the other hand, the new stringent experimental constraints in the plane $\delta$ versus $\theta_{23}$ seem to corner the STSO10 and indeed  a narrow range of values of $\delta$ is now requested.  If the errors will shrink around the current best fit value $\theta_{23} \simeq 47^{\circ}$ and $\delta \simeq -130^{\circ}$, STSO10 for $a_2 \lesssim 5$ would be ruled out. However, this could be accommodated  by values $\a_2\simeq 6$ (or even lower), a possibility that should be explored within specific models. In any case, even for such high values for $\a_2$, a favoured range $\d\simeq [-120^{\circ},-30^{\circ}]$ is confirmed. Therefore, future data, expected from long-baseline neutrino experiments, will  test the STSO10 solution in quite a  crucial way. In particular, the results from the anti-neutrino data expected from NO$\nu$A and more results from T2K
should help a more precise determination of $\theta_{23}$ and $\delta$.  A measurement of the effective neutrinoless double beta decay neutrino mass in the range $\simeq [10,30]\,{\rm meV}$, and a consequent deviation from normal hierarchy, would still provide an ultimate powerful test of the STSO10 solution.  

\vspace{-1mm}
\subsection*{Acknowledgments}

We acknowledge financial support from the STFC Consolidated Grant L000296/1. 
We also wish to thank Michele Re Fiorentin, Kareem Farrag and Teppei Katori for useful discussions.
Numerical calculations have been performed with the Iridis Computer Cluster at the University of Southampton. This project has received funding/support from the European Union’s Horizon 2020 research and innovation programme under the Marie Skłodowska-Curie grant agreement No 690575.

\end{document}